\documentclass[aps,prb,superscriptaddress,reprint,notitlepage]{revtex4-1}
\usepackage{amsmath}
\usepackage{graphicx}
\usepackage{bm}
\usepackage{float}
\usepackage{appendix}

\renewcommand{\vec}[1]{\mathbf{#1}}

\newcommand{\mctwo}{Department of Microtechnology and Nanoscience (MC2),
Chalmers University of Technology, SE-41296 Gothenburg, Sweden}

\begin{document}

\title{Extent of Fock-exchange mixing for a hybrid van der Waals density functional?}

\author{Yang Jiao}
\affiliation{\mctwo}
\author{Elsebeth Schr{\"o}der}
\affiliation{\mctwo}
\author{Per Hyldgaard}
\email[]{hyldgaar@chalmers.se}
\affiliation{\mctwo}

\date{\today}

\begin{abstract}
The vdW-DF-cx0 exchange-correlation hybrid design [J.\ Chem.\ Phys.\ 
{\bf 146},  234106 (2017)] has a truly nonlocal correlation component
and aims to facilitate concurrent descriptions of both covalent and
non-covalent molecular interactions. The vdW-DF-cx0 design mixes
a fixed ratio, $a$, of Fock exchange into the consistent-exchange 
van der Waals density functional, vdW-DF-cx [Phys.\ Rev.\ B 
{\bf 89}, 035412 (2014)]. The mixing value $a$ is sometimes taken as a 
semi-empirical parameter in hybrid formulations. Here, instead, we assert 
a plausible optimum average $a$ value for the vdW-DF-cx0 design from a 
formal analysis; A new, independent determination of the mixing $a$ is 
necessary since the Becke fit [J.\ Chem.\ Phys.\ {\bf 98}, 5648 (1993)], 
yielding $a'=0.2$, is restricted to semilocal correlation
and does not reflect non-covalent interactions.
To proceed, we adapt the so-called two-legged hybrid construction 
[Chem. Phys. Lett. {\bf 265}, 115 (1997)] to a starting point in the vdW-DF-cx 
functional. For our approach, termed vdW-DF-tlh, we estimate the properties 
of the adiabatic-connection specification of the exact exchange-correlation 
functional, by combining calculations of the Fock exchange and of the 
coupling-constant variation in vdW-DF-cx. 
We find that such vdW-DF-tlh hybrid constructions yield accurate characterizations 
of molecular interactions (even if they lack self-consistency). The accuracy
motivates trust in the vdW-DF-tlh determination of system-specific values of the Fock-exchange 
mixing. We find that an average value $a'=0.2$ best characterizes the vdW-DF-tlh 
description of covalent and non-covalent interactions, although there exists some 
scatter. This finding suggests that the original Becke value, $a'=0.2$, 
also represents an optimal average Fock-exchange mixing for the new, 
truly nonlocal-correlation hybrids. To enable self-consistent 
calculations, we furthermore define and test a zero-parameter hybrid 
functional vdW-DF-cx0p (having fixed mixing $a'=0.2$) and document that this 
truly nonlocal correlation hybrid works for general molecular interactions 
(at reference and at relaxed geometries). It is encouraging that  the 
vdW-DF-cx0p functional remains useful also for descriptions of some 
extended systems.

\end{abstract}

\maketitle

\section{Introduction}

An elegant and robust formulation of exchange-correlation (XC) hybrid functionals\cite{Becke93,PBE0,KimJor94,SteFri94,beckeperspective,DFcx02017} 
emerges by using the adiabatic-connection formula\cite{lape75,gulu76,lape77} 
(ACF) to balance exchange and correlation.\cite{Gorling93,Perdew96,Burke97,Ernzerhof97} The ACF establishes 
the XC energy as an integral of the electron-gas response dependence on 
the assumed strength $V_{\lambda}=\lambda V$ of the electron-electron 
interaction, $V$.  The $\lambda=0$ value is given by Fock exchange 
$E_{\rm x}^{\rm Fo}$. The ACF-based hybrid construction is 
relevant when the hybrid is based in plasmon- and constraint-based 
XC functionals,\cite{gulu76,lape77,lape80,lameprl1981,pewa86,pewa92,pebuer96,PBEsol,Dion,thonhauser,vv10,behy14,hybesc14,Thonhauser_2015:spin_signature,SCANvdW} where we can use a formal density-scaling analysis\cite{Levy85,Levy91,Levy95chapter} 
to reliably extract the nature of the electron-gas response at $\lambda\to 1$.
This density and coupling scaling analysis has been completed\cite{Perdew96,Burke97,Ernzerhof97,ScalingAnalysis} 
for both the semilocal PBE functional\cite{pebuer96,pebuwa96} and for truly 
nonlocal-correlation functionals of the van der Waals density functional (vdW-DF) method.\cite{Dion,Thonhauser_2015:spin_signature,Berland_2015:van_waals} The
scaling analysis for PBE leads to a rationale for the popular PBE0 hybrid,\cite{PBE0} computing $E_{\rm x}^{\rm Fo}$ from Kohn-Sham (KS) orbitals obtained in a self-consistent solution.

Some of us have recently extended the family of such ACF-based hybrids, launching 
nonlocal-correlation hybrid formulations, for example vdW-DF-cx0,\cite{DFcx02017}
based on the consistent-exchange vdW-DF-cx 
version.\cite{behy14,bearcoleluscthhy14} 
An exploration of vdW-DF-based hybrids
is motivated because the vdW-DF versions still have a GGA-type exchange and are 
thus prone to self-interaction errors.\cite{Becke93,beckeperspective,Perdew96} 
This limitation affects descriptions of charge-transfer processes in molecular
systems.\cite{JiaMen13,WaEsZe17} Also, the 
intra-molecular charge transfers affect non-covalent interactions between molecules.\cite{beloschy13,DFcx02017} The vdW-DF-cx0 hybrid is given by
\begin{equation}
E_{\rm xc}^{cx0}= a E_{\rm x}^{\rm Fo} + (1-a)E_{\rm x}^{\rm cx} + E_{\rm c}^{\rm cx} \, ,
\label{eq:cx0def}
\end{equation}
where $E_{\rm x(c)}$ denotes the exchange (correlation) component of vdW-DF-cx.
In launching the original vdW-DF-cx0 version, we picked a fixed Fock-exchange mixing
value $a=0.25$ in analogy with the construction of PBE0.\cite{PBE0,Perdew96} 
The $a=0.25$ choice is different from the $a'=0.2$
value that was originally suggested by 
Becke for molecular systems\cite{Becke93} and which is used, for example,
in the B3LYP hybrid.\cite{beckeperspective,KimJor94,SteFri94}

This paper seeks to answer two questions for the vdW-DF-cx0 design: (1) can we get away
with picking a single, all round, value of $a$ for the study of molecules, and, 
if so, (2) what would be a good mixing value $a$? The questions are important since
the vdW-DF-cx0 design aims to serve as a general purpose materials theory that can
deliver concurrent descriptions of both covalent and noncovalent
binding in molecules and in bulk. Our analysis is not based on the
full ACF-based hybrid construction\cite{Gorling93,Perdew96,Ernzerhof97} (using  
perturbation-theory studies to establish the $\lambda\to 0$ behaviors) for that 
would be prohibitively costly. Instead we pursue a bootstrap approach, using the 
so-called two-legged hybrid construction\cite{Burke97,Ernzerhof97} to 
define non-self-consistent (vdW-DF-cx-based) hybrids in a design called 
vdW-DF-tlh. Such constructions, summarized in Fig.\ \ref{fig:twolegO2} and below, 
are computationally much cheaper.
We simply have to use our previously-developed mapping of the coupling-constant 
scaling for the vdW-DF-cx functional,\cite{ScalingAnalysis} and adapt the original 
PBE-based analysis.\cite{Burke} Our vdW-DF-tlh construction can be cast in terms of 
Eq.\ (\ref{eq:cx0def}), however, with the key difference that the Fock-exchange mixing
$a_{\rm sys}$ is now explicitly asserted for each system and property of interest. From 
the computed $a_{\rm sys}$ values, we can answer our questions around an
optimum average mixing of Fock exchange in the vdW-DF-cx0 design.

We furthermore consider the question:  is there robustness in the vdW-DF-cx0 design? 
To answer this question, we define a zero-parameter (`0p') version, 
termed vdW-DF-cx0p (having Fock mixing $a'=0.2$ as motivated by the vdW-DF-tlh 
analysis). We contrast performance with that of vdW-DF-cx, the original vdW-DF-cx0 
(having Fock mixing $a=0.25$) version, in self-consistent, fully relaxed
calculations. We also compare the performance at reference geometries against 
that of dispersion-corrected GGA, meta-GGA, and against a traditional
(that is, semilocal-correlation) hybrid.\cite{gmtkn55}
Our test cases are molecule systems, subsets of the G2\cite{G2-97} and 
GMTKN55\cite{gmtkn55} benchmark sets, bulk semiconductors and a few transition 
metals.\cite{Csonka09} 

The rest of the paper is organized as follows. In Sec.\ II we present the
theory, summarizing the nature of the ACF-hybrid formulation  
and of the starting point, the consistent-exchange vdW-DF-cx version.\cite{behy14,bearcoleluscthhy14,hybesc14} 
Section III summarizes computational details and Sec.\ IV details 
the two-legged hybrid constructions, defining vdW-DF-tlh. 
Section V presents our vdW-DF-tlh analysis, discusses a plausible value for the 
Fock-mixing fraction in the vdW-DF-cx0 hybrid design, and presents a 
performance comparison. Finally, Sec.\ VI contains summary and conclusion.

\section{Theory}

Computing the density-density correlation function $\chi_\lambda$ at general values of the
coupling constant $\lambda$ for the electron-electron interaction $\lambda V$ permits a formally exact
determination of the XC energy, via the ACF\cite{lape75,gulu76,lape77}
\begin{equation}
E_{\rm xc} = - \int_0^\infty \, \frac{du}{2\pi} \, \hbox{Tr} \{ \chi_\lambda(iu) V \} -E_{\rm self}\,.
\label{eq:ACF}
\end{equation}
The last, so-called self-interaction term is just $E_{\rm self}= \hbox{Tr} \{ \hat{n} V\}/2$
where $\hat{n}$ denotes the density operator. For every $\lambda$ we can define an XC hole
\begin{equation}
n_{{\rm xc},\lambda}(\mathbf{r},\mathbf{r'}) = - \frac{1}{n(\vec{r})} \int_0^\infty \, \frac{du}{2\pi} \,
\chi_\lambda(\mathbf{r},\mathbf{r'}; iu) - \delta(\vec{r}-\vec{r'})
\label{eq:lambdahole}
\end{equation}
and an XC energy contribution
\begin{equation}
 E_{{\rm xc},\lambda}
 \equiv \frac{1}{2} \int_{\mathbf{r}}\int_{\mathbf{r'}}
\frac{n(\mathbf{r}) \, n_{{\rm xc},\lambda}(\mathbf{r},\mathbf{r'}) }{|\mathbf{r}-\mathbf{r}'|} \, .
\label{eq:ExcVar}
\end{equation}
The exact XC energy then results from a coupling constant integral
\begin{equation}
    E_{\rm xc}  =  \int_0^1 \, d\lambda \, E_{{\rm xc},\lambda} \, .
\end{equation}

\subsection{Consistent-exchange vdW-DF}

\begin{figure}
\includegraphics[width=0.45\textwidth]{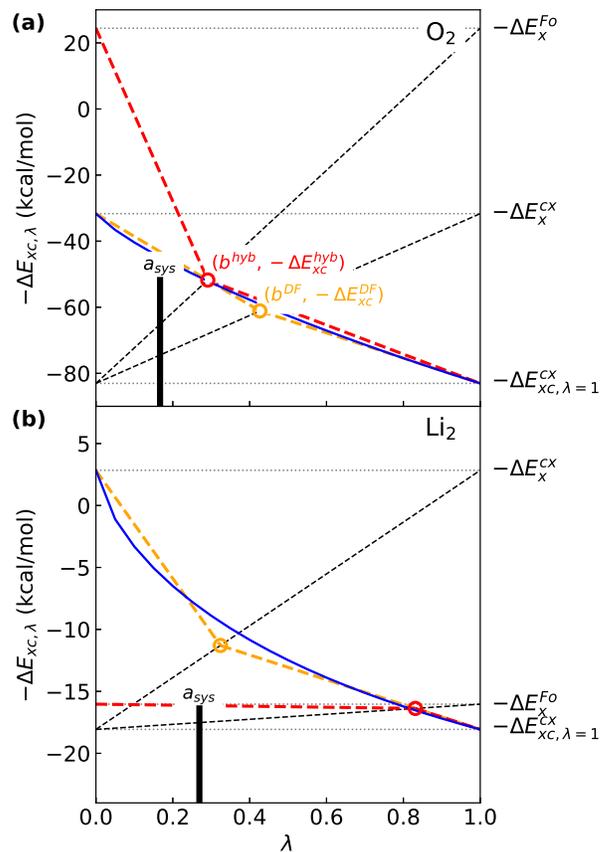}
\caption{Constructions of two-legged hybrids, termed vdW-DF-tlh, based on 
vdW-DF-cx. Here the vdW-DF-tlh approximation is used for analysis of the atomization energies of O$_2$ (top panel) and Li$_2$ (bottom panel). 
The panels show the two-legged 
representations (thick red dashed lines) of the $\lambda$-dependence of 
the exchange-correlation binding contributions.\cite{ScalingAnalysis} 
The solid blue curves show the $\lambda$-dependence of the vdW-DF-cx 
exchange-correlation binding contribution. The thin dashed lines are 
guiding the two-legged-hybrid construction.\cite{Burke97}
The orange dashed lines show the two-legged representations of this vdW-DF-cx 
variation, identifying the weighting $b^{\rm DF}$ (orange circles) between 
contributions evaluated at $\lambda\to 0$ and $\lambda\to 1$ 
limits of vdW-DF-cx. 
The red circles identify the weighting $b^{\rm hyb}$ of the $\lambda\to 0$ and 
$\lambda\to 1$ limits for a corresponding hybrid-vdW-DF-cx
construction. Finally, the pair of vertical thick bars identify the 
Fock-exchange mixing value $a_{\rm sys}$ that reflects the asserted value 
of $b^{\rm hyb}$.
}
\label{fig:twolegO2}
\end{figure}

The vdW-DF method\cite{rydberg03p126402,Dion,thonhauser,lee10p081101,hybesc14,behy14,Thonhauser_2015:spin_signature,Berland_2015:van_waals,DFcx02017}
is an attractive  framework for approximating the XC energy in density functional theory (DFT).
The method starts by considering the XC holes of a generalized gradient approximation 
(GGA) functional.\cite{Dion,thonhauser} It then adds a truly nonlocal correlation 
term $E_{\rm c}^{\rm nl}$  that systematically counts the total energy  
gain by the electrodynamic coupling between such semilocal XC holes.\cite{jerry65,ma,ra,anlalu96,hybesc14} 

The family of vdW-DF versions and variants\cite{Dion,dionerratum,lee10p081101,cooper10p161104,optx,vdwsolids,behy14,hamada14} 
permits computationally efficient\cite{roso09,libxcvdW,TraBla17} DFT studies of sparse 
materials,\cite{langrethjpcm2009} systems which have 
important regions with a low electron density. These truly nonlocal functionals
have by now found very broad applications, as summarized in 
Refs.~\onlinecite{langrethjpcm2009,rev8,Berland_2015:van_waals}. The same is true for the
related VV09 and VV10 functionals,\cite{vv09,vv10,Sabatini2013p041108} 
that use a different screening model for the account of nonlocal correlation effects.

In the vdW-DF method, we split the XC energy into a semilocal GGA-type functional
$E_{\rm xc}^0$ and a truly nonlocal-correlation term $E_c^{\rm nl}$. In general, there
is also a cross-over term $\delta E_{\rm x}^0$ related to the exchange description, 
as discussed elsewhere,\cite{behy14,hybesc14,bearcoleluscthhy14,Berland_2015:van_waals}
\begin{equation}
E_{\rm xc}^{\rm vdW-DF} = E_{\rm xc}^0 + E_{\rm c}^{\rm nl} + \delta E_{\rm x}^0 \, .
\label{eq:DFdef}
\end{equation}
The vdW-DF method can be interpreted as a computationally efficient evaluation of
the coupling-induced frequency shifts in the Ashcroft picture of 
vdW forces.\cite{jerry65,ma,ra,anlalu96,hybesc14} The long-range vdW forces are
described as arising from an electron-dynamical coupling between GGA-type XC holes 
and in the presence of the screening produced by the surrounding atoms.\cite{hybesc14}  

The recent consistent-exchange vdW-DF-cx version\cite{behy14} is crafted so
that it preserves current in its account of the electron-gas response.\cite{bearcoleluscthhy14} 
In practical terms, the consistent-exchange vdW-DF-cx formulation seeks to eliminate the adverse effects of the cross-over term $\delta E_{\rm x}^0$ in Eq.\ (\ref{eq:DFdef}), making it effectively an approximate mean-value evaluation\cite{behy14,hybesc14} of the ACF.\cite{lape75,gulu76,lape77} 
This is possible as long as the interaction is dominated by contributions with small 
values of the density gradient.\cite{berlandthesis,behy14,hybesc14,ScalingAnalysis} 
The vdW-DF-cx performs on par with or better than the popular 
GGAs\cite{pebuer96,PBEsol} for many bulk, surface, interface, and molecular 
properties.\cite{behy13,torbjorn14,ErhHylLin15,RaiPRB16,AmbSil16,HeBeBr16,LinErh16,LofErh16,RanPRB16,BrownAltvPRB16,Gharaee2017,KeSpMi17,Olsson17,BorArn17,CapAlo17,LonPopDes17,DFcx02017,Claudot18}  
It reliably accounts for van der Waals (vdW) forces in cases where interactions
compete,\cite{bearcoleluscthhy14,FriserSol16} for example, in the descriptions 
of weak chemisorption, oxide ferroelectrics, and metal-organic frameworks.\cite{bearcoleluscthhy14,Thonhauser_2015:spin_signature,Arnau16,WaZhKe17} 

\subsection{Coupling constant scaling and hybrids}

Use of hybrid XC functionals in DFT is in general
motivated by the observation that exchange dominates in several molecular
properties. A semilocal exchange form often leads to too much
confinement\cite{Becke93,beckeperspective} of the so-called XC
hole,\cite{gulu76,lape77} that reflects this electron-gas response.
The coupling-constant analysis of physically motivated functionals,\cite{Levy85,Levy91,Gorling93,Perdew96,Burke97,Ernzerhof97,ScalingAnalysis} permits us to pursue an ACF-based hybrid construction. 

The key observations are these. At physical conditions, corresponding 
to $\lambda=1$, the plasmons dominate the $\chi_{\lambda=1}$ 
behavior for homogeneous systems.\cite{pinesnozieres} The same is 
expected to hold in the weakly perturbed electron gas.\cite{lape80} 
Like the early formulations of the local density 
approximation (LDA),\cite{helujpc1971,gulu76,lape77} 
the consistent-exchange vdW-DF-cx explicitly emphasizes a plasmon 
foundation in its characterization of response ${\chi}_\lambda$ in the screened 
electron gas.\cite{behy13,bearcoleluscthhy14,Thonhauser_2015:spin_signature} 
This is done by crafting both exchange and correlation terms from a 
single-pole response model.\cite{hybesc14} A benefit is that we can 
expect $E_{{\rm xc},\lambda}^{\rm vdW-DF}$ to provide an 
accurate account in the $\lambda\to 1$ limit.
However, in the $\lambda\to 0$ limit the response $\chi_0$ must be different, 
given exclusively by single-particle excitation and exchange effects. 
In summary, we are motivated to extend the vdW-DF-cx design with a hybrid 
formulation, such as the recently formulated unscreened hybrid vdW-DF-cx0.\cite{DFcx02017} 

A more complete discussion is available by starting from the vdW-DF-cx
coupling-constant scaling.\cite{ScalingAnalysis} Here we just summarize the principle and the results that are relevant for the discussion of a hybrid vdW-DF-cx formulation.
The essential part is this result: using the simple density scaling
\begin{equation}
n(\mathbf{r}) \to n_{1/\lambda}(\mathbf{r}) \equiv n(\mathbf{r}/\lambda)/\lambda^3 \, ,
\label{eq:densscale}
\end{equation}
we can trace out the full coupling-constant variation using 
\begin{equation}
 E_{{\rm xc},\lambda}[n] = \frac{d}{d\lambda} \left\{ \lambda^2 E_{xc}[n_{1/\lambda}] \right\} \, .
\label{eq:scaleExc}
\end{equation}

The solid curves in Figure \ref{fig:twolegO2} show the corresponding coupling constant scaling as it emerges
for the XC energy contribution to the binding in the O$_2$ and Li$_2$ dimer.
The relevant quantity in typical DF theory calculations is energy differences, 
for example, the difference between molecule and atom total energies,
\begin{equation}
\Delta E^{\rm DF} = \sum_i E^{\rm DF}_{{\rm atom},i}-E^{\rm DF}_{\rm mol}\, .
\label{eq:EnDif}
\end{equation}
To discuss a hybrid formulation, we focus on the corresponding changes 
in the XC contributions $\Delta E_{\rm xc}[n]$ as well as on the differences 
that arise upon mapping the coupling constant scaling, $\Delta E^{\rm DF}_{{\rm xc},\lambda}$, 
and taking the $\lambda\to 1$ limit. For the hybrid vdW-DF constructions, we also need 
to compute binding-energy contributions, $\Delta E_{\rm x(c)}^{\rm DF}$, 
arising from the DF 
exchange (correlation) components.

The full ACF-based hybrid construction\cite{Gorling93} provides a formal argument 
that the Fock mixing in a hybrid construction (aiming to compute, for example, atomization energies) should be chosen in the form\cite{Perdew96,Burke97,Ernzerhof97}  
$a=1/m$, for $m=3,4,5,\ldots$.
The integer $m$ reflects the nature\cite{Perdew96} of the perturbation-theory calculation that enters in the full ACF-based hybrid construction.\cite{Gorling93} 
The value of $m$ and thus $a$ will, in principle, depend 
on both the system and the property that one wishes to 
investigate (as well as on the choice of underlying functional).  
On the other hand, the hybrid PBE0 and the hybrid vdW-DF-cx0, 
Eq.\ (\ref{eq:cx0def}), are only truly useful for making materials-specific predictions (of structure and binding) when they are kept deliberately free of parameters. Typically,  hybrid GGAs are used following the recommendation\cite{BurkePerspective} to stick with 
a fixed value of $a$ (one of the typical choices $a=0.25$ or $a=0.2$) since 
such choices are consistent with the formal nature of a full ACF-based hybrid 
construction.\cite{Perdew96} 

Here, we argue that the same approach should be used for vdW-DF-based hybrid constructions, including vdW-DF-cx0 that is based on vdW-DF-cx.\cite{DFcx02017}
Our discussion is based on crafting system-specific approximations to the full ACF-based hybrid construction,\cite{Gorling93,Perdew96,Ernzerhof97} adapting the ideas of the two-legged hybrid constructions for PBE.\cite{Burke97} Our approximation scheme, 
termed vdW-DF-tlh, permits us to discuss if a 
good (average) $a$ value can be found for using 
the vdW-DF-cx0 design\cite{DFcx02017} on problems defined by
general (covalent and non-covalent) interactions.

\section{Computational details}

\begin{table*}
\centering
\caption{Exchange-correlation contribution to atomization energies of molecules, in kcal/mol (1 eV = 23.06 kcal/mol). `cx' is short for 
vdW-DF-cx.  All results are  obtained for coordinates fixed in MP2(full)/6-31G(d) optimized geometries\cite{G2-97} and calculations are thus lower bounds on the atomization energies.  The test group is that used in Ref.\  \onlinecite{Burke97} to discuss the original PBE-based two-legged hybrid construction. The table also summarizes the performance in terms
of mean deviation (MD), mean absolute deviation (MAD), and mean absolute relative deviation (MARD) values.
%The systems are sorted into problems (first group) for which the two-legged hybrid construction is expected to work, %problems (second group) where there are strong static correlation, 
%making the construction fragile,\cite{Burke97} 
%and problems (second group) for which the %vdW-DF-cx-based 
%present construction differs qualitatively from %the PBE-based
%Ref.\ \onlinecite{Burke97} since $\Delta E_{\rm x}^{\rm HF} > \Delta E_{\rm x}^{\rm DF}$. 
\label{Tab:Ea}
 }
\begin{tabular*}{0.85\textwidth}{@{\extracolsep{\fill}}lrrrrrrrr}
\hline \hline
% Molecule & $\Delta E_x^{\rm HF}$ & $\Delta E_{\rm x}^{\rm cx}$ 
% & $\Delta E_{\rm xc}^{\rm cx}$ &   $\Delta E_{{\rm xc},\lambda=1}^{\rm cx}$   
% & $a$    &  $\Delta E^{cx}$  &  $\Delta E^{tlh}$  & $\Delta E^{ref}$\\
 Molecule & $\Delta E^{\rm ref}$ & $\Delta E^{\rm cx}$  & $\Delta E_x^{\rm HF}$
 & $\Delta E_{\rm x}^{\rm cx}$ &  $\Delta E_{\rm xc}^{\rm cx}$ 
 &   $\Delta E_{{\rm xc},\lambda=1}^{\rm cx}$   
 & $a_{\rm sys}$    &   $\Delta E^{\rm tlh}$ \\
\hline
          $\rm{LiH}$ &    58 &    58 &    31 &    28 &    51 &    64 &  0.13 &    58 \\
         $\rm{CH}_4$ &   420 &   429 &   168 &   222 &   307 &   362 &  0.21 &   418 \\
         $\rm{NH}_3$ &   298 &   303 &    82 &   127 &   202 &   252 &  0.21 &   293 \\
           $\rm{OH}$ &   106 &   105 &    31 &    28 &    55 &    73 &  0.17 &   106 \\
    $\rm{H}_2\rm{O}$ &   233 &   237 &    67 &   110 &   160 &   193 &  0.20 &   228 \\
           $\rm{HF}$ &   141 &   145 &    39 &    74 &    98 &   115 &  0.18 &   139 \\
         $\rm{Li}_2$ &    24 &    19 &    16 &    -3 &    11 &    18 &  0.27 &    24 \\
          $\rm{LiF}$ &   138 &   140 &   111 &   124 &   148 &   165 &  0.21 &   137 \\
  $\rm{C}_2\rm{H}_2$ &   406 &   417 &   112 &   237 &   310 &   361 &  0.17 &   395 \\
  $\rm{C}_2\rm{H}_4$ &   563 &   579 &   212 &   326 &   436 &   510 &  0.19 &   557 \\
          $\rm{HCN}$ &   313 &   320 &    11 &   125 &   188 &   232 &  0.17 &   301 \\
           $\rm{CO}$ &   259 &   265 &     5 &   101 &   137 &   164 &  0.15 &   251 \\
          $\rm{N}_2$ &   228 &   228 &   -96 &     6 &    59 &    97 &  0.17 &   211 \\
           $\rm{NO}$ &   153 &   161 &   -41 &    30 &    73 &   105 &  0.17 &   148 \\
         $\rm{Cl}_2$ &    58 &    70 &   -18 &    60 &    77 &    89 &  0.11 &    61 \\
          $\rm{O}_2$ &   120 &   138 &   -24 &    32 &    61 &    83 &  0.17 &   128 \\
          $\rm{H}_2$ &   110 &   112 &    23 &    30 &    57 &    73 &  0.22 &   110 \\
          $\rm{F}_2$ &    38 &    52 &   -95 &    10 &    27 &    41 &  0.10 &    42 \\
          $\rm{P}_2$ &   117 &   120 &   -68 &   -13 &    30 &    60 &  0.18 &   110 \\
\hline
Average  $a_{\rm sys}$ & \multicolumn{6}{c}{G2-1 subset}  &     0.18 &   \\
 MD(kcal/Mol)  &  &   5.95  &  &  &  &  &  & -3.45 \\ 
 MAD(kcal/Mol) &  &   6.68  &  &  &  &  &  &  5.02 \\ 
 MARD(\%)    &  &   6.47  &  &  &  &  &  &  2.89 \\ 
\hline
\end{tabular*}
\end{table*}

All of our calculations are based on the plane-wave \textsc{Quantum Espresso}
package \cite{QE,Giannozzi17} which has the consistent exchange vdW-DF-cx
version,\cite{behy14} as well as the rigorous spin extension
of the vdW-DF method.\cite{Thonhauser_2015:spin_signature} All molecule
studies use an 80~Ry wavefunction-energy cutoff. 
Core electrons are generally represented by Troullier-Martins type \cite{Troullier91p1993} norm-conserving pseudo potentials from 
the \textsc{abinit} package\cite{abinit05} in our studies of 
molecular properties. However, we used the \textsc{Quantum Espresso} 
projector augmented-wave (PAW) set up\cite{kjpaw} to also complete 
a set of PBE-XDM\cite{XDM07a,XDM12}
studies that we present in a performance comparison. 

For the vdW-DF-tlh  construction we rely on a numerical analysis of 
the coupling-constant scaling of density functional components, evaluating
Eq.\ (\ref{eq:scaleExc}). Here we use a post-processing code, termed 
\textsc{ppACF}, that we have described separately.\cite{ScalingAnalysis}  
As summarized below, we need only compute the differences 
in the $E_{{\rm xc},\lambda=1}$ values (as obtained for given densities). 

An  $8\times 8\times 8$ $k$-point sampling and optimized 
norm-conserving Vanderbilt (ONCV) pseudopotentials \cite{ONCV} 
are used for our vdW-DF-cx and vdW-DF-cx0(p) characterizations of bulk semiconductors 
and of a few transition metals.

We systematically rely on the adaptively compressed exchange (ACE)
operator\cite{linACE} for calculations of the Fock-exchange term $\Delta E_{\rm x}^{\rm Fo}$.
This is now a standard part of \textsc{Quantum Espresso},\cite{Giannozzi17} 
(although requiring a pre-compilation flag). Use of ACE speeds up hybrid calculations 
for molecules and it dramatically accelerates hybrid studies of extended systems.\cite{linACE} 
The ACE acceleration makes it possible to complete a self-consistent 
hybrid DFT calculation of a transition-metal element on the scale of 
hours (on a single, standard node of a high-performance computer).

\section{Hybrid construction: vdW-DF-tlh}

The hybrids rely on calculations of the Fock-exchange energy
\begin{equation}
E_{\rm x}^{\rm Fo}  =  - \frac{1}{2} \int_\mathbf{r} \int_{\mathbf{r'}} \frac{\tilde{n}_{1} (\mathbf{r},\mathbf{r'}) \tilde{n}_1(\mathbf{r'},\mathbf{r})}{|\mathbf{r}-\mathbf{r}'|} \, ,
\label{eq:foexA}
\end{equation}
where $n_1(\mathbf{r},\mathbf{r'}) =  \sum_i \phi^*_i(\mathbf{r})\phi_i(\mathbf{r'})$ and where $\phi_i$ denotes the set of 
solution wavefunctions. Fock-energy differences, $\Delta E_{\rm x}^{\rm Fo}$,
defined in analogy with Eq.\ (\ref{eq:EnDif}), are mixed with the exchange description of the underlying
density function (in our case vdW-DF-cx) to correct the description of, for example, band gaps and self-interaction effects.
For standard hybrid calculations the wavefunctions $\phi_i$ are taken as the KS solutions. For discussion of
atomization energies, however, we found that it was necessary to assert $\Delta E_{\rm x}^{\rm Fo}$
from self-consistent Hartree-Fock solutions.

\subsection{Two-legged hybrid approximation}

We first recall that any regular density functional `DF' should 
itself be seen as providing a cross-over between approximations for the 
single-particle and for the many-particle descriptions at $\lambda=0$ and 
$\lambda=1$, respectively.\cite{Burke97}  For the pure functional we simply inquire
when the $\lambda$ scaling of vdW-DF-cx, or of any functional `DF', 
intersects the opposite diagonal\cite{Burke97}
$[0,E_{{\rm xc},\lambda=1}^{\rm DF}]-[1,E_{\rm x}^{\rm DF}]$ at $[b^{\rm DF},E_{\rm xc}^{\rm DF}]$.
The intersection point,
\begin{equation}
 b^{\rm DF}[n]=\frac{E_{\rm xc}^{\rm DF}[n]-E_{{\rm xc},\lambda=1}^{\rm DF}[n]}{E_{\rm x}^{\rm DF}[n]-E_{{\rm xc},\lambda=1}^{\rm DF}[n]} \, ,
 \label{eq:bdensspec}
\end{equation}
determines the weighting of $\lambda\to0$ and $\lambda \to 1$ components
\begin{equation}
E^{\rm DF}_{\rm xc}[n] = b^{\rm DF}[n] E_{\rm x}^{\rm DF}[n]
+ (1- b^{\rm DF}[n]) E_{{\rm xc},\lambda=1}^{\rm DF}[n] \,.
\label{eq:mixingDFfunct}
\end{equation}
Note that $b^{\rm DF}[n]$ is itself a functional of the density -- but that is just part of the overall 
`DF' description. For a description of `DF' energy differences,
\begin{equation}
\Delta E^{\rm DF}_{\rm xc} = b^{\rm DF}_{\rm sys} \Delta E_{\rm x}^{\rm DF}
+ (1- b^{\rm DF}_{\rm sys}) \Delta E_{{\rm xc},\lambda=1}^{\rm DF} \,,
\label{eq:mixingDFsystem}
\end{equation} 
there is, consequently, an explicit system dependence on the weighting, 
$b^{\rm DF}_{\rm sys}$, of $\lambda\to0$ and $\lambda\to 1$ contributions, 
as indicated by the subscript `sys'.

The hybrid vdW-DF-cx constructions should be seen as a natural generalization
of the regular-functional mixing behavior Eq.\ (\ref{eq:mixingDFsystem}); the generalization
is motivated by the fact that the Fock-exchange differences $\Delta E_{\rm x}^{\rm Fo}$, 
Eq.\ (\ref{eq:foexA}), are generally more accurate than the DF exchange description. The two-legged non-empirical hybrid constructions \cite{Perdew96,Burke97} 
use $\Delta E_{{\rm x}}^{\rm Fo}$ to anchor the $\lambda\to 0$ limit, 
expressing a corrected weighting
\begin{equation}
 \Delta E_{\rm xc}^{\rm hyb}= b^{\rm hyb}_{\rm sys} 
 \Delta E_{\rm x}^{\rm Fo}+(1-b^{\rm hyb}_{\rm sys}) \Delta E_{\rm xc,\lambda=1}^{\rm DF} \, . 
 \label{eq:Hybridweight}
\end{equation}
By establishing the new weighting factors $b^{\rm hyb}_{\rm sys}\neq b^{\rm DF}_{\rm sys}$, below,
we also determine rational choices for the mixing of a Fock exchange term as expressed
in a more common hybrid construction form
\begin{equation}
 \Delta E_{\rm xc}^{\rm hyb}=a_{\rm sys} \Delta E_{\rm x}^{\rm Fo}+(1-a_{\rm sys}) \Delta E_{\rm x}^{\rm DF} +
E_{\rm c}^{\rm DF} \, .
\label{eq:hybmix}
\end{equation}
The formal relation to Eq.\ (\ref{eq:Hybridweight}) is given by
\begin{equation}
 a_{\rm sys}=\frac{\Delta E_{\rm xc}^{\rm hyb}-
 \Delta E_{\rm xc}^{\rm DF}}{\Delta E_{\rm x}^{\rm Fo}-\Delta E_{\rm x}^{\rm DF}} \, .
 \label{eq:asyseval}
\end{equation}

For the two legged constructions\cite{Burke97} we define and approximate two gradients
\begin{align}
 g^{\rm L}_{\rm sys} & = \left. \frac{d \Delta E_{{\rm xc},\lambda}}{d\lambda}\right|_{\lambda=0} 
   \approx - \frac{\Delta E_{\rm x}^{\rm DF}-\Delta E_{\rm xc}^{\rm DF}}{b^{\rm DF}_{\rm sys}} \, , \\
 g^{\rm R}_{\rm sys} & = \left. \frac{d\Delta E_{{\rm xc},\lambda}}{d\lambda}\right|_{\lambda=1} 
   \approx - \frac{\Delta E_{\rm xc}^{\rm DF}- \Delta E_{{\rm xc},\lambda=1}^{\rm DF}}
   {1-b^{\rm DF}_{\rm sys}} \, .
\end{align}
The first gradient should ideally be computed in perturbation theory, leading to the non-empirical 
hybrid construction discussed in Ref.\ \onlinecite{Perdew96}; An exploration of this 
approach is beyond the present scope. Both gradients are instead approximated, as 
indicated, by a linear form given by $b_{\rm sys}^{\rm DF}$, namely, the 
position of the kink in a two-legged construction for vdW-DF-cx itself.
 
Besides the vdW-DF-cx energy differences, we also compute
the Fock exchange energy difference to provide a vdW-DF-cx- and system-specific 
determination of $g^{\rm L}_{\rm sys}$ (and $g^{\rm R}_{\rm sys}$ when relevant). Adapting the logic of Ref.\ \onlinecite{Burke97}, the value of 
$g^{\rm L}_{\rm sys}$ specifies a motivated approximation for 
balancing $\lambda\to 0$ and $\lambda\to1$ 
contributions in cases where $\Delta E_{\rm x}^{\rm Fo} < 
\Delta E_{\rm x}^{\rm DF}$. The value of $g^{\rm R}_{\rm sys}$
is only relevant for cases where $\Delta E_{\rm x}^{\rm Fo} >
\Delta E_{\rm x}^{\rm DF}$ and its use requires an additional 
discussion, below. 

The top panel of Fig.~\ref{fig:twolegO2} illustrates the vdW-DF-tlh construction 
for a typical molecular-binding case, binding in the O$_2$ molecule. 
The orange circle identifies the
crossing point between the vdW-DF-cx coupling-constant scaling 
(blue curve) and the diagonal 
(lower black dotted line) from 
$(\lambda=0, -\Delta E_{{\rm xc},\lambda=1})$ to $(\lambda=1, 
- \Delta E_{\rm x}^{\rm DF})$.
The crossing point $(\lambda=b_{\rm sys}^{\rm DF}, -\Delta E^{\rm DF}_{\rm xc})$ is system and 
property specific, with the value of $b_{\rm sys}^{\rm DF}$ given by the generalization of
Eq.\ (\ref{eq:bdensspec}) to energy differences. The dashed orange curve shows a two-legged
approximation for the actual DF coupling constant variation; this curve has a kink at
the orange circle. 

Figure \ref{fig:twolegO2} summarizes our vdW-DF-tlh constructions. The figure
shows that there are differences between the Fock 
and DF exchange binding contributions, $\Delta E_{\rm x}^{\rm Fo}$ and 
$\Delta E_{\rm x}^{\rm DF}$. We seek revised coupling-constant curves that better
approximate the coupling-constant variation in the ACF determination of 
the exact XC functional. The $\Delta E_{\rm x}^{\rm Fo}-\Delta E_{\rm x}^{\rm DF}$
difference is used to define two-legged constructions,\cite{Burke97} 
for example, the red dashed curves in Fig.\ \ref{fig:twolegO2}, 
anchored by $- \Delta E_{\rm x}^{\rm Fo}$ in the $\lambda \to 0$ limit and 
the trusted $ - \Delta E_{{\rm xc},\lambda=1}$ value at the other end.
In essence, we first use Eq.\ (\ref{eq:Hybridweight}) to 
complete the vdW-DF-tlh calculation of the energy difference and we then use 
Eq.~(\ref{eq:hybmix}) 
to extract the mixing $a_{\rm sys}$ that is equivalent to this vdW-DF-tlh description.

Our two-legged hybrid constructions seek to follow the DF coupling-constant scaling 
as far as possible as we move to lower coupling constant values.\cite{Burke97} This 
leads to placing the kink (red circle) in the revised two-legged curve (dashed red curve)
both on the vdW-DF-cx coupling constant curve and on the second indicated diagonal (upper black dotted line).
The intersection, or kink (red circle) identifies the plausible value of a revised weighting\cite{Burke97}
\begin{equation}
% b^{\rm hyb,L}_{\rm sys}=\frac{\Delta E_{\rm x}^{\rm DF}- \Delta E_{{\rm xc},\lambda=1}^{\rm DF}}
%          {\Delta E_{\rm x}^{\rm Fo}-\Delta E_{{\rm xc},\lambda=1}^{\rm DF} + g^{\rm L}_{\rm sys}} \, .
b^{\rm hyb,L}_{\rm sys}=\frac{\Delta E_{\rm x}^{\rm DF}- \Delta E_{{\rm xc},\lambda=1}^{\rm DF}}
          {\Delta E_{\rm x}^{\rm Fo}-\Delta E_{{\rm xc},\lambda=1}^{\rm DF} - g^{\rm L}_{\rm sys}} \,  .
        \label{eq:Lweight}  
\end{equation}
of $\lambda\to 0$ and $\lambda \to 1$ limits.
In this case, the revised two-legged approximation suggests that the
plausible coupling-constant curve would also be downward concave. The
$\lambda$ value $b^{\rm hyb,L}_{\rm sys}$ will be located left of 
$b_{\rm sys}^{\rm DF}$, as identified by the superscript. The $b^{\rm hyb,L}_{\rm sys}$
is finally converted into a system (and property) specific value of a plausible 
Fock-exchange mixing value $a_{\rm sys}$, given by Eq.\ (\ref{eq:asyseval}), and as
identified by the thick vertical bar.

\begin{table*}
\centering
\caption{Summary of two-legged hybrid construction and comparison of performance for 
the 55 molecule atomization energies of the G2-1 data set,\cite{G2-97} for 17 molecular reaction 
energies in the G2RC data set,\cite{gmtkn55} and for 26 ionization potentials in the
G21IP data set.\cite{gmtkn55} All energies are in kcal/mol (1 eV = 23.06 kcal/mol). 
'cx' is short for vdW-DF-cx. The results are
evaluated at coordinates fixed in MP2(full)/6-31G(d) optimized geometries for
G2-1\onlinecite{G2-97} and at experimental geometries for G2RC and for G21IP.\cite{gmtkn55} 
The Fock exchange terms are calculated using the orbitals of a 
Hartree-Fock (vdW-DF-cx) calculations for G2-1 (for G2RC and for G21IP).
The average mixing ratios, $a_{\rm sys}$, are 0.18, 0.21, and 0.18 respectively.
\label{Tab:Ea-G2-1}
}
\begin{tabular*}{0.85\textwidth}{@{\extracolsep{\fill}}lrrrrr}
\hline \hline
 Molecule  &  $\Delta E^{\rm PBE0}$   & $\Delta E^{\rm cx}$ & $\Delta E^{\rm tlh}$ 
           & $\Delta E^{\rm cx0}$ & $\Delta E^{\rm cx0p}$ \\
\hline
 G2-1 atomization & $a=0.25$ & $a=0$ & $\langle a_{\rm sys} \rangle = 0.18$ & $a=0.25$ & $a'=0.20$ \\
 %MD(kcal/mol)  &  -4.37 [-3.36] &  8.28 [9.09] &      -1.46 &   -2.85 [-1.48] &  -0.61 [ 0.04]
 MD(kcal/mol)  &  -4.37 &  8.28 &      -1.46 &   -2.85 &  -0.61 \\
 %MAD(kcal/mol) &   5.50 [ 4.99] &  8.94 [9.73] &       4.29 &    5.58 [ 4.90] &   4.18 [ 4.13]
 MAD(kcal/mol) &   5.50 &  8.94 &       4.29 &    5.58 &   4.18 \\ 
 %MARD(\%)      &   4.42 [ 4.12] &  6.47 [7.36] &       2.98 &    4.17 [ 3.68] &   3.24 [ 3.27]
 MARD(\%)      &   4.42 &  6.47  &       2.98 &    4.17 &   3.24 \\
\hline
 G2RC subset (17 reactions) & $a=0.25$ & $a=0$ & $\langle a_{\rm sys} \rangle = 0.21 $ & $a=0.25$ & $a'=0.20$\\
MD (kcal/mol)   &  -3.80  &  -0.10  &  -1.80  &  -2.34  &  -1.89  \\
MAD (kcal/mol)  &   5.68  &   5.78  &   4.90  &   4.43  &   4.54  \\
MARD (\%)       &  38.03  &  57.08  &  38.55  &  35.10  &  38.46  \\
\hline
 G21IP subset (26 molecules) & $a=0.25$ & $a=0$ & $\langle a_{\rm sys} \rangle = 0.18$ & $a=0.25$ & $a'=0.20$\\
 MD(kcal/mol)  & -1.22  &   -2.67  &   -3.43  &  -0.02  & -0.32   \\
 MAD(kcal/mol) &  5.27  &    4.25  &    5.57  &   4.19  &  4.22   \\
 MARD(\%)      &  2.16  &    1.66  &    2.28  &   1.75  &  1.76   \\
\hline
\end{tabular*}
\end{table*}

The lower panel shows how we have adapted the two-legged constructions for
descriptions of the atomization energies for Li$_2$, LiH, and OH, and for
other cases where $E_{\rm x}^{\rm Fo}>E_{\rm x}^{\rm DF}$.
Aiming again to align the coupling constant scaling behavior in the 
large-$\lambda$ limit, we then place the kink (red circle) of the revised two-legged approximation (red dashed line)
at $(b_{\rm sys}^{{\rm hyb},R},\Delta E_{\rm xc}^{\rm DF})$, where 
\begin{equation}
 b^{{\rm hyb},R}_{\rm sys}=
%  \frac{g^{\rm R}_{\rm sys}}
%    { \Delta E_{\rm x}^{\rm Fo}-\Delta E_{{\rm xc},\lambda=1}^{\rm DF}+g^{\rm L}_{\rm sys}} \, .
  \frac{-g^{\rm R}_{\rm sys}}
    { \Delta E_{\rm x}^{\rm Fo}-\Delta E_{{\rm xc},\lambda=1}^{\rm DF}-g^{\rm R}_{\rm sys}} \, .
    \label{eq:Rweight}
\end{equation}
That is, in such adjusted two-legged constructions, 
$b^{{\rm hyb},R}_{\rm sys}$ is the $\lambda$ value that 
formally specifies the weighting of $\lambda\to 0$ and 
$\lambda\to 1$ limits. As indicated
by the superscript `$R$',  we then have 
$b^{{\rm hyb},R}_{\rm sys} > b_{\rm sys}^{\rm DF}$,
leading to larger values of the corresponding Fock-exchange 
mixing value $a_{\rm sys}$. 

Table \ref{Tab:Ea} summarizes the vdW-DF-tlh constructions, 
showing the $a_{\rm sys}$ and atomization-energy results. 
The table focuses on the systems that were originally analyzed 
for two-legged hybrid constructions based on PBE.\cite{Burke97} We find that there 
is a spread in the predicted values of $a_{\rm sys}$. We also find that the binding energies 
$\Delta E^{\rm tlh}$ that are predicted with vdW-DF-tlh, improve the description relative to that
provided by the vdW-DF-cx starting point. However, it is important to point out that the vdW-DF-tlh constructions are introduced for analysis purposes only. 

\subsection{Limitations of vdW-DF-tlh usage}

There are four fundamental and practical problems with using vdW-DF-tlh.
First, it cannot be cast in a self-consistent formulation and thus cannot be used
for general materials characterizations, nor will it always be accurate.\cite{Burke97} 
Second, the design logic breaks down completely when $-\Delta E_{\rm x}^{\rm Fo}$ 
is lower than   $-\Delta E_{{\rm xc},\lambda}^{\rm DF}$. Third, it
is not
clear that the two-legged construction holds for cases where it implies 
using a small value of the Fock mixing ratio, $a_{\rm sys}< 0.15$, in Eq.\ (\ref{eq:hybmix}), for reasons discussed
in Refs.\ \onlinecite{Perdew96,Burke97,Ernzerhof97}. 
Fourth, in cases where $E_{\rm x}^{\rm Fo} < E_{\rm x}^{\rm DF}$ (for example, 
as in the bottom panel of Fig.\ \ref{fig:twolegO2})
it is not easy to motivate the particular vdW-DF-tlh description as a
plausible approximation to the exact ACF XC-functional specification\cite{gulu76,lape77} or even to the full ACF-based hybrid construction.\cite{Gorling93}

The last point deserves an additional discussion as it impacts our 
analysis. The coupling constant variation in $E_{{\rm xc},\lambda}$
should be downward concave in the exact ACF evaluation.\cite{Levy85,Levy91,Levy95chapter} 
This downward-concave behavior is correctly reflected in 
the coupling-constant variations that represent the vdW-DF-cx functional behavior.
With a full ACF-based hybrid construction\cite{Gorling93} we
would also expect a downward concave coupling constant variation, having a form
similar to that shown by the red dashed line in the top panel of Fig.\ \ref{fig:twolegO2}. 

However, the approximate, non-self-consistent vdW-DF-tlh construction 
sometimes produces an 
upward concave coupling-constant variation, bottom panel of Fig.\ \ref{fig:twolegO2}.
Such variations, for the specific vdW-DF-tlh constructions, are not 
motivated by formal theory,\cite{Levy91,Levy95chapter} even if these 
vdW-DF-tlh constructions are themselves fairly accurate in characterizations
of atomization energies, molecular-reaction energies, and 
ionization potential energies (supplementary materials 
Tables S.I, S.III, and S.IV, respectively). Interestingly, the 
issue does not appear in the vdW-DF-tlh description of our set
of non-covalent inter-molecular bonding, supplementary materials, 
Table S.V.

We interpret this practical problem for vdW-DF-tlh usage for characterization
of standard molecular properties (atomization, reaction, and ionization
energies) as a warning of insufficient accuracy in the two-legged constructions. 
We note that vdW-DF-tlh is not self-consistent so we need to have a plausible guess 
for the character of the wavefunctions when computing the Fock-exchange 
term. In hybrid-DFT calculations, we generally use KS wavefunctions (here
obtained in self-consistent vdW-DF-cx calculations) but these are not directly 
relevant for a characterization of the Fock-exchange energies of atom and molecules. 
They will not always produce $\Delta E_{\rm x}^{\rm Fo}$ estimates that are  accurate.

Supplementary materials Table S.II shows a vdW-DF-tlh characterization of atomization
energies, when instead we use self-consistent Hartree-Fock calculations to determine the Fock-exchange energy. The table shows that this adjustment in the two-legged hybrid  construction ensures that a down-ward concave variation underpins the vdW-DF-tlh analysis for atomization energies for all but four cases. One of these exceptions is the case of the Li dimer, the specific example that we analyze in the bottom panel of Fig.\ \ref{fig:twolegO2}. Computing $\Delta E_{\rm x}^{\rm Fo}$ from Hartree-Fock solution wavefunctions is, on the other hand, not motivated for the study of reaction or of general ionization energies.

We base our analysis of the vdW-DF-cx0 design, below, on vdW-DF-tlh 
constructions in which $\Delta E_{\rm x}^{\rm Fo}$ is computed 
from vdW-DF-cx wavefunctions for all but the atomization energies
(where we use HF). This means including cases
with an upward concave scaling behavior, for example, in the case of
reaction and ionization energies (Tables S.II-IV). The impact on 
our analysis and for the predictions for an average 
mixing value $a_{\rm sys}$, however, is limited. For molecular 
atomization, reaction, and ionization energies (Tables S.II-IV), the 
set of vdW-DF-tlh constructions suggest an average  Fock mixing value $a = \langle 
a_{\rm sys} \rangle = 0.186 \approx 0.2$ for the vdW-DF-cx0 design.\cite{DFcx02017} 
If instead we had restricted the analysis set to cases with a resulting downward-concave coupling-constant variation,\cite{Levy91,Levy95chapter,Burke97} 
the average would be $\langle a_{\rm sys} \rangle = 0.178$.

\begin{figure}
\includegraphics[width=0.45\textwidth]{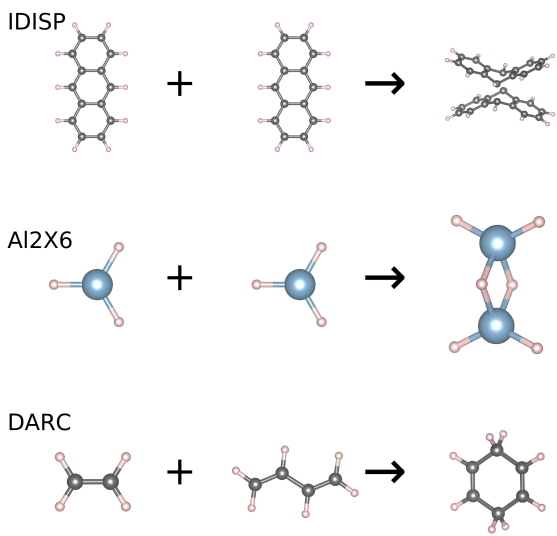}
\caption{Tests of the role of charge relocation in general molecular interaction: representative examples from
the benchmark sets\cite{gmtkn55} of intra-molecular noncovalent binding
(IDISP, top row), of aluminum dimerization 
(Al2X6, middle row), and of Diels-Alder reaction
energies (DARC, bottom row).\cite{gmtkn55}
}
\label{fig:schemIntra}
\end{figure}

\section{Results and discussions}

\begin{table*}    
\caption{Binding energies of the S22 dataset. The geometries are optimized at either the CCSD(T) or MP2 level as taken from Ref.~\onlinecite{S22-06}. 
The reference interaction energies are taken from Ref.~\onlinecite{S22-11} as suggested by the
GMTKN55 data set.
%The reference interaction energies are all calculated using both CCSD(T)/CBS counterpoise corrected (CP) and MP2/CBS CP as taken from Ref.~\onlinecite{S22-10}.
The Fock exchange term is calculated using vdW-DF-cx orbitals.}
\label{tab:S22}
\begin{tabular*}{0.85\textwidth}{@{\extracolsep{\fill}}lrrrrrr}
\hline \hline
     & $\Delta E^{\rm ref}$ & $\Delta E^{\rm cx}$ & $\Delta E^{\rm tlh}$ & $\Delta E^{\rm cx0}$   & $\Delta E^{\rm cx0p}$  &  $a_{\rm sys}$ \\    
\hline
                           Ammonia dimer &   3.133 &    2.63 &    2.79 &    2.87 &    2.82 &    0.20 \\
                             Water dimer &   4.989 &    4.57 &    4.77 &    4.86 &    4.80 &    0.21 \\
                       Formic acid dimer &  18.753 &   18.65 &   19.38 &   19.64 &   19.42 &    0.21 \\
                         Formamide dimer &  16.062 &   14.93 &   15.72 &   15.91 &   15.71 &    0.21 \\
                   Uracil dimer h-bonded &  20.641 &   19.01 &   19.78 &   20.04 &   19.82 &    0.20 \\
    2-pyridoxine 2-aminopyridine complex &  16.934 &   16.88 &   17.13 &   17.32 &   17.22 &    0.18 \\
    Adenine-thymine Watson-Crick complex &  16.660 &   15.45 &   15.82 &   16.06 &   15.92 &    0.18 \\
\hline
 Average  $a {\rm sys}$       & \multicolumn{5}{c}{Hydrogen Bonding}  &     0.20   \\
 MD(kcal/Mol)    &  &    -0.72 &    -0.26 &   -0.07 &   -0.21 &  \\
 MAD(kcal/Mol)   &  &     0.72 &     0.49 &    0.43 &    0.48 &  \\
 MARD(\%)        &  &     6.78 &     4.44 &    3.63 &    4.26 &  \\
\hline
                           Methane dimer &   0.527 &    0.63 &    0.75 &    0.79 &    0.76 &    0.19 \\
                            Ethene dimer &   1.472 &    0.98 &    1.33 &    1.42 &    1.33 &    0.20 \\
                 Benzene-methane complex &   1.448 &    1.29 &    1.55 &    1.65 &    1.58 &    0.19 \\
        Benzene dimer parallel displaced &   2.654 &    2.60 &    3.04 &    3.25 &    3.12 &    0.17 \\
                          Pyrazine dimer &   4.255 &    3.90 &    4.44 &    4.69 &    4.53 &    0.18 \\
                      Uracil dimer stack &   9.805 &    9.30 &   10.40 &   10.79 &   10.49 &    0.19 \\
            Indole-benzene complex stack &   4.524 &    4.27 &    4.88 &    5.18 &    5.00 &    0.17 \\
           Adenine-thymine complex stack &  11.730 &   10.84 &   12.15 &   12.69 &   12.31 &    0.18 \\
\hline
 Average $a {\rm sys}$ & \multicolumn{5}{c}{Dispersion Bonding}  &     0.18   \\
 MD(kcal/Mol)    &  &    -0.33 &     0.27 &    0.51 &    0.34 &  \\
 MAD(kcal/Mol)   &  &     0.35 &     0.30 &    0.52 &    0.37 &  \\
 MARD(\%)        &  &    11.60 &    11.98 &   16.53 &   13.54 &  \\
\hline
                   Ethene-ethyne complex &   1.496 &    1.54 &    1.69 &    1.75 &    1.70 &    0.19 \\
                   Benzene-water complex &   3.275 &    2.94 &    3.30 &    3.41 &    3.32 &    0.20 \\
                 Benzene-ammonia complex &   2.312 &    2.06 &    2.36 &    2.46 &    2.38 &    0.19 \\
                     Benzene-HCN complex &   4.541 &    4.11 &    4.69 &    4.83 &    4.68 &    0.21 \\
                  Benzene dimer T-shaped &   2.717 &    2.55 &    2.93 &    3.09 &    2.98 &    0.18 \\
          Indole-benzene T-shape complex &   5.627 &    5.24 &    5.80 &    6.01 &    5.86 &    0.18 \\
                            Phenol dimer &   7.097 &    6.29 &    6.92 &    7.14 &    6.97 &    0.19 \\
\hline
 Average   $a {\rm sys}$      &  \multicolumn{5}{c}{Mixed Bonding}  &     0.19   \\
 MD(kcal/Mol)    &  &    -0.33 &     0.09 &    0.23 &    0.12 &  \\
 MAD(kcal/Mol)   &  &     0.35 &     0.14 &    0.23 &    0.15 &  \\
 MARD(\%)        &  &     8.28 &     4.71 &    7.88 &    5.27 &  \\
\hline
\hline
 Average  $a_{\rm sys}$       & \multicolumn{5}{c}{All S22 dimers}  &     0.19   \\
 MD(kcal/Mol)    &  &    -0.45 &     0.04 &    0.24 &    0.09 &  \\
 MAD(kcal/Mol)   &  &     0.47 &     0.31 &    0.40 &    0.34 &  \\
 MARD(\%)        &  &     9.01 &     7.27 &    9.68 &    7.95 &  \\
\hline
\end{tabular*}
\end{table*}

The usefulness of our two-legged hybrid vdW-DF-cx constructions is that 
they provide analysis of the nature of and the extent of exchange mixing 
$a=1/m$ in the vdW-DF-cx design.\cite{DFcx02017} We note that the 
formal ACF-based hybrid construction\cite{Gorling93,Perdew96,Ernzerhof97} motivates that we should pick $m=4$, $m=5$, or perhaps $m=6$.\footnote{Fock-exchange mixing values  $a=1/m$ for $m > 6$ is not considered relevant\cite{Burke97}} 

A necessary condition for using our vdW-DF-tlh constructions for an analysis of the vdW-DF-cx0 design is that the 
vdW-DF-tlh characterizations can be seen as accurate. In the following we therefore list a summary of the vdW-DF-tlh constructions
and of the asserted average mixing values $\langle a_{\rm sys} \rangle$, 
in concert with statistical analysis of performance for molecular interactions, as compared  against quantum-chemistry reference calculations.\cite{G2-97,G2RC,gmtkn55} 

\begin{table*}
\caption{Two-legged hybrid construction and comparison of 
performance for systems with pronounced
intra-molecular noncovalent binding (IDISP),
for covalent binding in aluminum complexes (Al2X6), and for 
the Diels-Alder set of reaction energies (DARC). 
The geometries and reference energies for the IDISP (6 systems),
Al2X6 (6 systems), and DARC (14 systems) data sets are taken from Ref.~\onlinecite{gmtkn55}.
The values in square bracket are calculated with optimized geometries using 
the corresponding functionals -- in the case of IDISP: including/excluding the
C$_{22}$H$_{46}$ unfolding case (the only case where there is any discernible
relaxation effect, see Table S.VI-VIII).
\label{tab:intramol}
}
\begin{tabular*}{0.85\textwidth}{@{\extracolsep{\fill}}lrrrrr}
\hline \hline
Reaction   & $\Delta E^{\rm PBE0}$ &
             $\Delta E^{\rm cx}$ &
             $\Delta E^{\rm tlh}$ &
             $\Delta E^{\rm cx0}$ &
             $\Delta E^{\rm cx0p}$ \\
\hline
IDISP             & $a=0.25$ & $a=0$  & $\langle a_{\rm sys} \rangle = 0.18 $ & $a=0.25$ & $a'=0.20$ \\   
MD (kcal/mol)     & 1.42    &  2.02[2.73/2.67]  &  1.01  &  0.78[1.30/0.75]  &  1.03[1.56/1.11]  \\
MAD (kcal/mol)    & 9.77    &  2.36[2.73/2.67]  &  1.66  &  1.90[2.40/2.07]  &  1.75[2.32/2.03]  \\
MARD (\%)         & 241.98  & 31.52[50.64/29.97] & 22.44  & 28.79[53.38/22.88] & 24.00[51.11/22.59]  \\
\hline
Al2X6             & $a=0.25$ & $a=0$  & $\langle a_{\rm sys} \rangle = 0.20 $ & $a=0.25$ & $a'=0.20$ \\   
MD(kcal/mol)      & -3.29[-3.32] & -2.67[ -2.67] & -1.70 &  -1.65[ -1.67] & -1.86[ -1.88] \\
MAD(kcal/mol)     & 3.29[3.32]  &  2.67[  2.67] &  1.70 &   1.65[  1.67] &  1.86[  1.88] \\
MARD(\%)          & 10.05[10.16] &  6.81[  6.81] &  4.51 &   4.21[  4.26] &  4.75[  4.81] \\
\hline
DARC              & $a=0.25$ & $a=0$  & $\langle a_{\rm sys} \rangle = 0.21 $ & $a=0.25$ & $a'=0.20$ \\   
MD (kcal/mol)     &  1.06[1.08] & -0.84[-0.86] & -4.00 & -4.74[-4.72] & -3.94[-3.94] \\   
MAD (kcal/mol)    &  3.28[3.28] &  1.70[1.70] &   4.00 &  4.74[4.72]  &  3.94[3.94] \\   
MARD (\%)         & 12.80[12.82] &  5.60[5.52] &  12.49 & 15.39[15.35] & 12.30[12.33] \\   
\hline
\end{tabular*}
\end{table*}

Overall, we present and summarize vdW-DF-tlh constructions for atomization energies 
(the G2-1 set, supplementary materials Table S.II), for reaction energies 
(subset of G2RC, Table S.III), for ionization energies (subset of G21IP, 
Table S.IV), for inter-molecular non-covalent interactions 
(S22, Table S.V), for intra-molecular non-covalent interactions (IDISP, 
Table S.VI), for binding energies of aluminum dimers (Al2X6, S.VII), and 
for a set of Diels-Alder reaction energies (DARC, Table S.VIII), 
using reference geometries and reference binding energies in the 
G2\cite{G2-97} and the GMTKN55\cite{gmtkn55} benchmark sets.  
For each of the benchmark (and construction) subsets, 
we report the vdW-DF-tlh specification of averaged 
Fock-exchange mixing parameter $\langle a_{\rm sys} \rangle$
and the mean deviation (MD), mean absolute deviation (MAD)
and mean absolute relative deviations (MARD) values that 
characterize the vdW-DF-tlh constructions.

Figure \ref{fig:schemIntra} shows representative examples of the IDISP,
Al2X6, and DARC benchmark sets.  We include these benchmark sets because 
they are expected to reflect effects of the electron affinity and 
delocalization errors on both non-covalent and covalent molecular 
binding, and thus challenge hybrid formulations.\cite{G2RC} The 
intra-molecular non-covalent IDISP set has 
a low average absolute relative energy of about 14 
kcal/mol\cite{gmtkn55} and represents cases where the 
weaker vdW interaction competes with other binding
mechanisms.\cite{bearcoleluscthhy14,behy13} The DARC set are defined
from systems involving also double and triple bonds. Here one might expect 
better performance from a meta-GGA (like SCAN\cite{SCAN} or
dispersion-corrected versions thereof\cite{SCANvdW,gmtkn55})
than from a standard hybrid.\cite{G2RC} 
They are important tests on our analysis of and search for a plausible
average value of the Fock-exchange mixing parameter $\langle a_{\rm sys} 
\rangle$ in the vdW-DF-cx0 design.\cite{DFcx02017} They allow us to test
if the average $\langle a_{\rm sys} \rangle\approx 0.2$, obtained 
by the vdW-DF-tlh analysis of atomization, reaction, 
and ionization cases (discussed in Sec.\ IV) also
holds in more general molecular interaction cases.

\begin{table*}
\centering
\caption{
Performance of vdW-DF-cx and vdW-DF-cx0p for S22 data set compared with dispersion corrected DFT 
methods: PBE-XDM,\cite{XDM07a,XDM12} PBE-D3,\cite{grimme3} and PBE-TS.\cite{ts09}
The geometries are taken from Ref.~\onlinecite{S22-06} and the reference energies are taken from Ref.~\onlinecite{S22-11}.
The values in brackets are from fully relaxed calculations. 
In the lower part is the statistics of the binding distance for fully relaxed dimers.
The PBE-XDM results are obtained  using \texttt{QUANTUM ESPRESSO} with PAW pseudopotentials.\cite{kjpaw}
\label{tab:s22vdwcorr}
}
\begin{tabular*}{0.85\textwidth}{@{\extracolsep{\fill}}lrrrrr}
\hline \hline
	    & $\Delta E^{\rm cx}$   &
	      $\Delta E^{\rm cx0p}$ &
	      $\Delta E^{\rm PBE-XDM}$ &
	      $\Delta E^{\rm PBE-D3}$ &
	      $\Delta E^{\rm PBE-TS}$ \\
\hline
	MD (kcal/mol)  & -0.45[-0.60] &  0.09[-0.31] & -0.41[-0.36] & -0.09[-0.20] &  0.20[0.02]  \\
	MAD (kcal/mol) &  0.47[0.68]  &  0.34[0.60]  &  0.59[0.54]  &  0.54[0.57]  &  0.34[0.49]  \\
	MARD (\%)      &  9.01[9.43]  &  7.95[9.90]  &  9.75[8.40]  & 11.56[11.05] & 10.01[12.83] \\
\hline \hline
	   & $d^{\rm cx}-d^{\rm ref}$   &
	     $d^{\rm cx0p}-d^{\rm ref}$ &
	     $d^{\rm PBE-XDM}-d^{\rm ref}$ &
	     $d^{\rm PBE-D3}-d^{\rm ref}$ &
	     $d^{\rm PBE-TS}-d^{\rm ref}$ \\
\hline
MD (\AA)   & 0.058 & 0.034 & 0.064 & 0.061 & 0.018 \\
MAD (\AA)  & 0.068 & 0.044 & 0.075 & 0.073 & 0.040 \\
MARD (\%)  & 2.38 & 1.57 & 2.65 & 2.57 & 1.50 \\
\hline
\end{tabular*}
\end{table*}

\subsection{Summary of vdW-DF-tlh constructions and performance}

Table \ref{Tab:Ea-G2-1} reports an overview of the performance of 
vdW-DF-tlh, compared with that of PBE0 and of vdW-DF-cx itself, 
and as obtained  for a range of benchmarks 
for covalent molecular binding.  The table 
also lists the average of the $a_{\rm sys}$ values 
that results in the two-legged constructions. The distribution of
such $a_{\rm sys}$ values centers on $\langle a_{\rm sys} \rangle=0.2$.
However, there is also some scatter in the vdW-DF-tlh
specification of $a_{\rm sys}$ values, as seen in
supplementary materials tables S.II and  S.III.

Table \ref{Tab:Ea-G2-1} shows that, overall, the performance of 
vdW-DF-tlh is good for covalent molecular binding properties, as compared 
with PBE0, with vdW-DF-cx, and with the original vdW-DF-cx0 version.\cite{DFcx02017} 
To provide a fair comparison the table lists PBE0, vdW-DF-cx, and vdW-DF-cx0 
results for energies at the reference geometries that were also
used in the vdW-DF-tlh characterizations; we shall return to a discussion
of relaxation effects below. The MD, the MAD, and the MARD 
decrease for the set of G2-1 atomization energies. The 
vdW-DF-tlh remains comparable to  the performance of PBE0 and 
vdW-DF-cx0 for the subset of reaction energies. The vdW-DF-tlh performance 
for ionization energies seems to slightly worsen, however, likely because 
the absence of a self-consistent determination affects the orbital description.

Table \ref{tab:S22} reports a summary of the two-legged construction and a 
comparison of  the vdW-DF-tlh performance for the S22 benchmark set,\cite{S22-06,S22-10} focusing on inter-molecular binding energies. 
The table is divided into cases with hydrogen, dispersion, and mixing 
bonding cases, but also summarizes the overall performance as compared 
to vdW-DF-cx and to vdW-DF-cx0, Ref.\ \onlinecite{DFcx02017}.
Again, there is some scatter in the predicted plausible $a_{\rm sys}$ values
but the average is centered on 0.2. The performance of vdW-DF-tlh is better than 
that of vdW-DF-cx0 in that the good account of hydrogen bonding is preserved while vdW-DF-tlh avoids some of the errors that vdW-DF-cx0 makes for purely vdW bonded cases.\cite{DFcx02017}

Table \ref{tab:intramol} shows a summary of the vdW-DF-tlh construction 
as well as a performance comparison for the IDISP, Al2X and DARC 
benchmark sets.\cite{G2RC} There is for these additional vdW-DF-tlh construction 
cases only a small scatter of the $a_{\rm sys}$ values around the average value $0.20$, 
as detailed in the supplementary materials.  

The description for IDISP is good already at the level of vdW-DF-cx (as further discussed below) and improves with the hybrid formulations. 
The two-legged hybrid constructions (`vdW-DF-tlh') is better than the original vdW-DF-cx0 version and approaches that of 
PBE0-D3, Ref.\ \onlinecite{gmtkn55}. 
The same trend is found also for the Al2X6 set. The deviations
for vdW-DF-cx and hybrids are larger
but so is the average absolute energy and the MARD values
are smaller than in the case of IDISP. Again the hybrid
vdW-DF formulations, including vdW-DF-tlh, improve the description.

Previous studies have indicated that meta-GGA descriptions (with dispersion 
corrections as in SCAN-D3) perform better than at least traditional 
semilocal hybrids for the DARC set.\cite{G2RC,gmtkn55} 
Here the regular vdW-DF-cx functional\cite{behy14} performs 
very well. Use of the hybrid vdW-DF formulations 
worsens this performance although
the two-legged construction (`vdW-DF-tlh') still
performs at the level of PBE0 (Table \ref{tab:intramol}).

We trust the vdW-DF-tlh analysis, summarized in Table \ref{tab:intramol}, for 
all of the IDISP, Al2X6, and DARC sets. Our trust builds on the observation 
that vdW-DF-cx, the original vdW-DF-cx0 version,\cite{DFcx02017} and a new 
zero-parameter (`0p') specification, denoted vdW-DF-cx0p and defined below, 
are all highly accurate in their characterization of structure. This is true
for all of the Al2X6 and DARC systems and all but one of the IDISP systems.  

Supplementary material Tables S.VII-VIII report 
results obtained at reference geometries as well as (inside square brackets)
at fully relaxed geometries. There are no discernible structural relaxations in
either of cx, cx0 (or cx0p) and consequently no relevant
energy differences in the binding and reaction energies that we have 
obtained for Al2X6 and DARC. 

\begin{figure}
\includegraphics[width=0.45\textwidth]{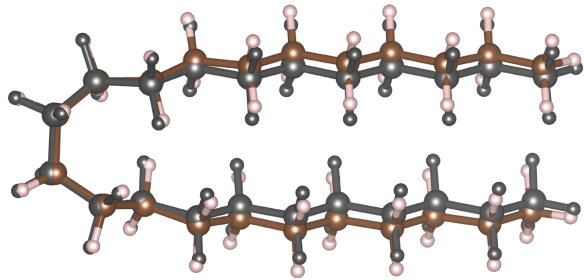}
\caption{Schematics of the folded C$_{22}$H$_{46}$ system
as described at the GMTKN55 reference geometries \cite{gmtkn55}
(black) and at the geometry relaxed in vdW-DF-cx calculations (brown 
and white). This structure defines the last of the IDISP benchmarks
and it is the only case of all of the IDISP, Al2X6, and DARC systems
where there are any discernible relaxations in vdW-DF-cx, vdW-DF-cx0,
and vdW-DF-cx0p calculations (relative to the quantum-chemistry reference 
data on structure \cite{gmtkn55}).
}
\label{fig:schemUnfold}
\end{figure}

Figure \ref{fig:schemUnfold} compares the structure result of a 
fully relaxed vdW-DF-cx calculation against that of the
quantum-chemistry reference data \cite{gmtkn55} for 
the folded morphology of C$_{22}$H$_{46}$; There are similar
structure differences for vdW-DF-cx0 and vdW-DF-cx0p calculations.
Such results are relevant for the last of the IDISP benchmark tests,
namely concerning C$_{22}$H$_{46}$ unfolding.  Supplementary material 
Table S.VI documents that there are no observable relaxation effect in 
any of the the first 5 benchmark cases of the IDISP benchmark set (as 
in all of the Al2X6 and DARC cases). 

The total relaxation effect on the results for the C$_{22}$H$_{46}$ unfolding energy is 
found to be just 2-3 kcal/mol for the hybrid vdW-DF versions, Table S.VI. However, the 
relaxation does cause a change of sign for all of the vdW-DF-cx, vdW-DF-cx0, and vdW-DF-cx0p 
results (relative to the quantum-chemistry reference data \cite{gmtkn55})
for this particular case. The relaxation effects translate into a 
finite impact on the IDISP MARD value, Table \ref{tab:intramol}.
At the same time, the MARD increase with relaxations is still 
originating from just one structure, Fig.\ \ref{fig:schemUnfold}.

\subsection{Definition of a parameter-free hybrid: vdW-DF-cx0p}

\begin{table*}
\caption{
Comparison of performance of vdW-DF-cx (`cx'), vdW-DF-cx0 (`cx0') and vdW-DF-cx0p (`cx0p')
against that of dispersion-corrected functionals, PBE-D3, revPBE-D3, SCAN-D3, and PBE0-D3.
The comparison is made for subsets of the GMTKN55 benchmark database.\cite{gmtkn55} The 
table reports mean absolute deviation (MAD) values in kcal/mol for calculations performed at 
reference geometries and against reference energies listed in Ref.\ \onlinecite{gmtkn55}.
The statistics for the dispersion-corrected functionals are taken from Ref.~\onlinecite{gmtkn55}
and are computed in an orbital-based approach, not the plane-wave pseudopotential approach that we
use here for the vdW-DF versions.
\label{tab:D3-MAD}
}
\begin{tabular*}{0.85\textwidth}{@{\extracolsep{\fill}}lrrr|rrrr}
\hline \hline
Reaction   & $\Delta E^{\rm cx}$ &
             $\Delta E^{\rm cx0}$ &
             $\Delta E^{\rm cx0p}$ &
             $\Delta E^{\rm PBE-D3}$ &
             $\Delta E^{\rm revPBE-D3}$ &
             $\Delta E^{\rm SCAN-D3}$ &
             $\Delta E^{\rm PBE0-D3}$ 
             \\
\hline
G2RC     & 6.16 & 4.01 & 4.24 & 6.92 & 6.16 & 6.39 & 6.75 \\
G21IP    & 4.08 & 4.13 & 4.10 & 3.84 & 4.20 & 4.69 & 3.68 \\
S22      & 0.47 & 0.40 & 0.34 & 0.48 & 0.43 & 0.47 & 0.48 \\
IDISP    & 2.36 & 1.90 & 1.75 & 2.76 & 3.14 & 2.05 & 1.54 \\
Al2X6    & 2.67 & 1.65 & 1.86 & 1.63 & 2.07 & 2.13 & 1.48 \\
DARC     & 1.70 & 4.74 & 3.94 & 3.31 & 3.71 & 2.01 & 3.76 \\
\hline
\end{tabular*}
\end{table*}

The vdW-DF-tlh analysis of the vdW-DF-cx0 design\cite{DFcx02017}
motivates us to define a new version,
\begin{equation}
E_{\rm xc}^{\rm cx0p}= a' E_{\rm x}^{\rm Fo} + (1-a')E_{\rm x}^{\rm cx} + E_{\rm c}^{\rm cx} \, ,
\label{eq:cx0primedef}
\end{equation}
to enable self-consistent calculations for general molecular (covalent and 
noncovalent) interactions. The choice of the $a'=0.2$ value represents an optimal 
average value for general molecular interactions. A motivation for introducing this 
vdW-DF-cx0p version, Eq.\ (\ref{eq:cx0primedef}), is that it can be seen as strictly free
of adjustable parameters. It is parameter free in the sense that the mixing value $a'$ 
is asserted from a formal analysis (i.e., our vdW-DF-tlh constructions) that reflects the expected ACF behavior of vdW-DF-cx.\cite{ScalingAnalysis}
 
At the same time, it should be made clear that the two-legged construction 
vdW-DF-tlh clearly identifies some scatter in the set of suggested 
$a_{\rm sys}$ mixing values (even if concentrated around $\langle a_{\rm sys} \rangle = 0.2$).
On the one hand, the scatter in $a_{\rm sys}$ could just be a consequence of 
vdW-DF-tlh not being self consistent. On the other hand, the scatter does suggest 
that the simple, unscreened hybrid vdW-DF-cx0 design, and the vdW-DF-cx0p version in 
particular, cannot be expected to be accurate for all types of molecular problems, let 
alone bulk-systems cases.  

To provide insight on the relevance of either of these possibilities, we are led to next
explore the performance of the vdW-DF-cx0p version for molecular and some extended systems.
A simple comparison of the vdW-DF-cx0p usefulness with that of vdW-DF-cx,\cite{behy14} of 
vdW-DF-tlh (above), of the original vdW-DF-cx0 version,\cite{DFcx02017} and with that of 
other hybrids or vdW inclusive functionals can provide some guide lines for how to best 
continue the development of truly nonlocal-correlation hybrids like the vdW-DF-cx0 
design. Also, it may be that while there are some scatter in the vdW-DF-tlh assessment 
of plausible $a_{\rm sys}$ values, there may not be a large impact of changing the 
Fock-exchange mixing in the vdW-DF-cx0 design. That is, the vdW-DF-cx0p version might 
provide a good all-round description of covalent and noncovalent
interaction properties in any case.

\begin{figure}
\includegraphics[width=0.45\textwidth]{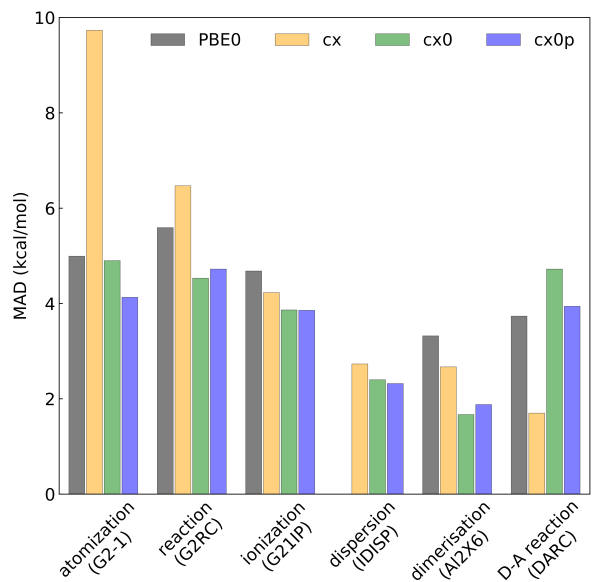}
\caption{Comparison of PBE0, vdW-DF-cx (cx), vdW-DF-cx0 (cx0), and vdW-DF-cx0p (cx0p) performance for binding properties 
of small molecules, for the G2-1 subsets of the G2 set\cite{G2-97} and for the G2RC, G21IP, IDISP, Al2X6, and DARC subsets 
of the GMTKN55 benchmark set.\cite{gmtkn55}  We compare the mean-average deviations that reflect fully relaxed results, 
measured against the reference energy data.
}
\label{fig:molMAD}
\end{figure}

\subsection{Robustness of the vdW-DF-cx0p: molecular tests}

We first note that tables \ref{Tab:Ea-G2-1}, \ref{tab:S22} and \ref{tab:intramol} also include a 
raw summary of the vdW-DF-cx0p performance comparison, for fixed geometries. Additional 
details for individual systems in the full G2-1, G2RC, G21IP, IDISP, Al2X6, DARC, and S22 benchmark 
sets are given in the supplementary material, Tables S.VI-S.XII. We find that vdW-DF-cx0p performs on 
par with vdW-DF-tlh for both covalent and non-covalent binding properties, even if slightly worse in the 
case of atomization and dispersion energies. Moreover, vdW-DF-cx0p works on par with or improves the 
vdW-DF-cx0 performance for covalent binding properties. The vdW-DF-cx0p functional also improves the 
description of dispersion and mixed binding cases of S22, although slightly worsening the description 
of cases with a pronounced hydrogen bond. 

\begin{figure*}
\includegraphics[width=0.7\textwidth]{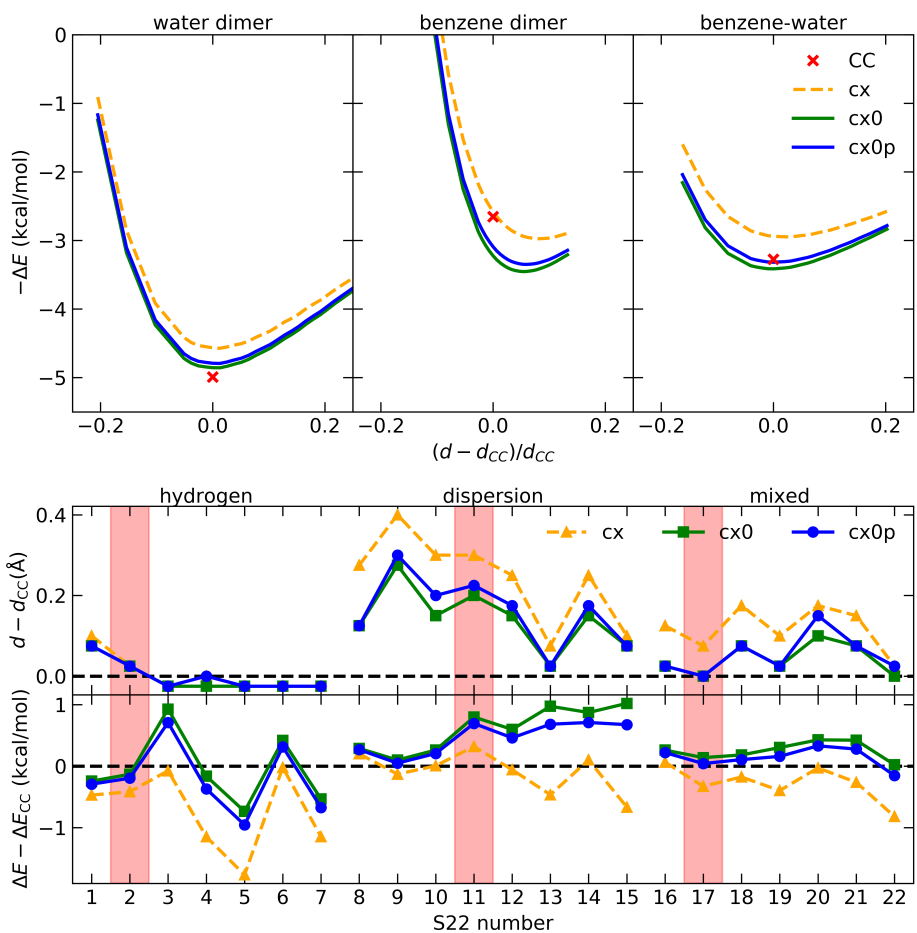}
\caption{Comparison of performance of vdW-DF-cx, of vdW-DF-cx0 and of vdW-DF-cx0p, for (inter-molecular) noncovalent binding in the S22 
data set.\cite{S22-06,S22-10} The top panel illustrates how, in this performance comparison, we partly include the effects of geometry relaxations, computing (for each functional) 
binding-energy curves as we decrease or increase the inter-molecular 
separation relative to the S22 reference (marked by a cross). 
The bottom panels compare the functional-specific results for the binding separation $d$ and the binding energy $\Delta E$, respectively. The variation is plotted relative to the S22 reference values, $d_{\rm CC}$ and $\Delta E_{\rm CC}$.
}
\label{fig:revS22}
\end{figure*}

Table \ref{tab:s22vdwcorr} shows a comparison of the vdW-DF-cx and vdW-DF-cx0p performance for the S22 data set against so-called dispersion-corrected GGA descriptions: PBE-XDM,\cite{XDM07a,XDM12} PBE-D3,\cite{grimme3} and PBE-TS.\cite{ts09}  
These are descriptions in which a pair-potential formulation of the vdW attraction is added to PBE. The numbers outside square parenthesis reflect results obtained at 
reference geometries and we find that performance of vdW-DF-cx and vdW-DF-cx0p functionals, as measured in the MAD values, are comparable to those of the dispersion-corrected versions. This is true in spite of the fact that the group of dispersion-corrected functionals (PBE-XDM,
PBE-D3, PBE-TS) employ a damping function which is fitted to training sets that themselves include (or consist of) the S22 data set. The damping in the PBE-XDM and PBE-D3 
versions is detailed, for example, in Ref.\ \onlinecite{gmtkn55} and is broader than
S22, comprising about 60 systems. For assessment of the S22 performance of vdW- or dispersion-corrected functionals, there are documented effects of focusing the
training of the damping function on just the S22 set.\cite{XDM12} Our vdW-DF-cx 
and vdW-DF-cx0p functionals avoid such parameter issues completely, being set 
by formal many-body physics input\cite{thonhauser,hybesc14} (including 
the ACF-based argument for picking the $a'=0.2$ mixing).

Table \ref{tab:D3-MAD} compares the MAD values that we
here obtain for vdW-DF-cx0p (`cx0p') against those obtained 
in PBE-D3, revPBE-D3, SCAN-D3 and PBE0-D3 in the GMTKN55
report,\cite{gmtkn55} and against the results that we obtain
in vdW-DF-cx and in the original vdW-DF-cx0 version.\cite{DFcx02017}
Ref.\ \onlinecite{gmtkn55} highlights (rev)PBE-D3 and SCAN-D3  
as good, all round choices on the lower rungs of functional approximations
and PBE0-D3 represents a natural hybrid reference (being a 
vdW-inclusive extension of a popular semilocal-correlation hybrid, PBE0).
The comparison is made for the here-investigated GMTKN55 subsets
and at listed reference geometries. Taken together, the set of benchmarks
probes the ability of vdW-DF-cx and vdW-DF-cx0p to describe a range of
molecular-interaction properties. As such, Table \ref{tab:D3-MAD}
can be seen as a supplement to Ref.\ \onlinecite{Claudot18}, which 
found that vdW-DF-cx performs significantly better for a
broad comparison of molecule problems than both PBE-TS\cite{ts09} 
and the so-called TS-MBD extension.\cite{ts-mbd} 

We can only provide a qualitative discussion since, in contrast to our  
present cx0p/cx/cx0 plane-wave calculations, the 
set of dispersion-corrected results are obtained in 
quantum-chemistry codes using quadruple-$\zeta$
atomic orbitals.\cite{gmtkn55} Nevertheless, we 
can observe that the vdW-DF-cx0p and vdW-DF-cx0 have the
best performance for reaction energies. On average, 
the vdW-DF-cx and vdW-DF-cx0p perform the same level
as the listed set of dispersion-corrected functionals.

Figure \ref{fig:molMAD} summarizes a comparison of PBE0, vdW-DF-cx, vdW-DF-cx0, 
and vdW-DF-cx0p performance for covalent, intra-molecular binding, 
using the aforementioned subsets of G2 and GMKTKN55\cite{G2-97,gmtkn55} but allowing 
for full structural approximations. The underlying data is given in the 
supplementary material, Tables S.VI-XI.  The zero-parameter
hybrid vdW-DF-cx0p performs better overall than PBE0 and vdW-DF-cx0 
for these G2 and GMTKN55 subsets but the improvements are just moderate.

Next we contrast the performances for G2-1, G2RC, and G21IP in 
fully-relaxed hybrid studies, Fig.\ \ref{fig:molMAD}, with those obtained 
using reference geometries, Table \ref{Tab:Ea-G2-1}. We find 
that both vdW-DF-cx0 and vdW-DF-cx0p characterizations for 
relaxed geometries are clearly better (essentially unchanged) for G2-1 
(for G2RC and G21IP). This is somewhat in contrast to the behavior for 
vdW-DF-cx itself. For PBE0 there is only limited effects of including
relaxations for G2-1, G2RC, and G21IP. 

As already discussed in subsection V.A and as detailed in Table
\ref{tab:intramol}, we find that vdW-DF-cx, vdW-DF-cx0, and vdW-DF-cx0p all 
perform excellently for structure characterizations in the IDISP, Al2X6, and 
DARC sets.  This is encouraging for the vdW-DF-cx and hybrid vdW-DF-cx formulations since 
these cases are included in the GMTKN55 \cite{gmtkn55} to test 
the ability of functionals to describe the effect of intra-molecular 
charge relocation on weaker binding energies.\cite{gmtkn55}

Fig. \ref{fig:schemUnfold} shows the structure of the folded
C$_{22}$H$_{46}$ system. As mentioned above, this is the one case where 
the vdW-DF-cx and hybrid vdW-DF-cx structure results differ from the reference
data among the IDISP/Al2X6/DARC benchmarks (supplementary materials 
Tables S.VI-VIII). We note that the IDISP benchmark subset is itself 
small,\cite{gmtkn55} and it is relevant to assert the extent that this
single difference affects our vdW-DF-cx/vdW-DF-cx0p benchmarking
for the IDISP set. Accordingly, we include in Table \ref{tab:intramol} 
a summary of benchmarking with fully-relaxed IDISP results while both
including and excluding this special C$_{22}$H$_{46}$ unfolding case; If the special case
is omitted we find again a good vdW-DF-cx and vdW-DF-cx0p performance
for IDISP, even when characterized in terms of the MARD values.

The comparison of vdW-DF-cx and vdW-DF-cx0 performance for the
DARC deserves a separate discussion. The DARC set probes 
description of systems with multiple bonds and this is a class of 
systems where the vdW-DF-cx already performs very well, Fig.\ 
\ref{fig:molMAD}. However, comparing the vdW-DF-cx 
MD and MAD values for the DARC set in Table \ref{tab:intramol}, it 
is also clear that vdW-DF-cx provides a systematic underestimation 
of the DARC energies. Since, furthermore the use of vdW-DF-cx0p 
systematically decreases these reaction energies, we end up with a 
worse performance for vdW-DF-cx0p. Our finding that the DARC set 
challenges vdW-DF-cx0p is consistent with previous findings 
that semilocal-correlation hybrids perform worse than, for example, 
meta-GGA based descriptions for the DARC set.\cite{G2RC} 

In the case of non-covalent inter-molecular bonding there is 
a potential for more pronounced geometry relaxations (not reflected in Table \ref{tab:S22}) to play a role. For example, in the S22 set of molecular dimers, the binding energies are small and the geometries can be significantly adjusted by forces. Computing binding energies at a fixed geometry need not provide a relevant description of the binding energy minima as described with a given (hybrid) functional. Thus ignoring relaxation might prevent us from learning if the functional in question is strongly over-binding. 

The top panels of Fig.\ \ref{fig:revS22} illustrate the procedure that we provide for a revised comparison of the performance for S22, going beyond the Table II comparison. 
In this we follow the approach that we have also previously used,  for example, in the recent paper launching the vdW-DF-cx0 hybrid vdW-DF design.\cite{DFcx02017} We begin with  reference dimer geometries,
which in the S22 benchmark case are  extracted from a series 
of increasingly more accurate quantum chemistry calculations, 
ending with an interpolation among coupled cluster (CC) 
studies for geometries located around the expected 
energy minimum. We next adjust the inter-molecular 
separations inwards and outwards in steps of 0.025 {\AA}
and obtain a binding curve for vdW-DF-cx, for vdW-DF-cx0, and for vdW-DF-cx0p. Finally, we extract both the molecular-dimer binding 
separation $d$ and binding energy estimates $\Delta E$ relative to the
S22 reference data, $d_{\rm CC}$ and $\Delta E_{\rm CC}$. 

The top panels also reveal the importance of including structural
relaxations. While a comparison at the $d_{\rm CC}$ geometry 
works well (for regular and hybrid vdW-DF-cx) in the case of 
hydrogen bonded systems (like the water dimer) and in the case of mixed 
bonding (like the water-benzene complex), it leads to an underestimation 
of the binding energy in systems (like the benzene-benzene complex)
with a pure dispersion bonding. This is one good reason to use 
vdW-DF-cx0p instead of vdW-DF-tlh.

The bottom panel of Figure \ref{fig:revS22} shows the full comparison of 
performance for vdW-DF-cx, vdW-DF-cx0, and 
vdW-DF-cx0p; the underlying data is listed in the supplementary 
materials, Table S.XII. The upper (lower) half of this panel reports the
variation in $d-d_{\rm CC}$ (in $\Delta E - \Delta E_{\rm CC}$).  We 
have (as in Table III) separated the survey into cases reflecting 
hydrogen binding, dispersion bonding, and mixed bonding. As also observed in 
Ref.\ \onlinecite{DFcx02017}, we find that hybrid vdW-DF-cx (vdW-DF-cx0 
and vdW-DF-cx0p) is accurate for hydrogen and mixed bonding cases 
but less accurate for the cases with a pure dispersion bonding.

\begin{table}[h]
\caption{Comparison of vdW-DF-cx (abbreviated `cx'), vdW-DF-cx0 (`cx0'), and 
vdW-DF-cx0p (`cx0p') performance for bulk semiconductors and a few second-row 
transition metals: cubic-cell lattice constant $b$ (in {\AA}), cohesive energy 
$E_{\rm coh}$ (in eV), and bulk modulus $B$ (in GPa).
Reference energies are experimental values corrected (except for $B$) by an 
estimate for zero-point energy corrections, as listed in
Ref.\ \onlinecite{Csonka09}.}
\label{tab:solids}
\begin{tabular}{llcccccr}
\hline
\hline
   &  & PBE & PBE0 &
             cx &
            cx0 &
             cx0p &
             Ref. \\
\hline

C	 
& $b$ 
& 3.572 & 3.553
& 3.561 & 3.550
& 3.552
& 3.543
\\
& $E_{\rm coh}$ 
& 7.671 & 7.505
& 7.841 & 7.572 
& 7.618
& 7.583
\\
& $B$ 
& 430 & 463
& 440 & 467
& 463
& 443
\\
Si	 
& $b$ 
& 5.464 & 5.441
& 5.437 & 5.430
& 5.431
& 5.416
\\
& $E_{\rm coh}$
& 4.504 & 4.558
& 4.743 & 4.723
& 4.766
& 4.681
\\
& $B$ 
& 89 & 99
& 93 & 101
& 100
& 99
\\
SiC	 
& $b$ 
& 4.374 & 3.352
& 4.358 & 4.346
& 4.348 
& 4.342 
\\
& $E_{\rm coh}$ 
& 6.368 & 6.353
& 6.590 & 6.489
& 6.542
& 6.488
\\
& $B$ 
& 210 & 229
& 217 & 232
& 229
& 225
\\
GaAs	 
& $b$ 
& 5.745 & 5.645
& 5.680 & 5.608
& 5.622 
& 5.638 
\\
& $E_{\rm coh}$ 
& 3.130 & 3.191
& 3.408 & 3.411
& 3.407
& 3.393
\\
& $B$ 
& 60 & 76
& 67 & 80
& 77
& 76
\\
\hline
Rh	 
& $b$ 
& 3.837 & 3.791
& 3.789 & 3.764
& 3.768 
& 3.793 
\\
& $E_{\rm coh}$ 
& 5.939 & 4.111
& 6.407 & 4.496
& 4.814
& 5.784
\\
& $B$ 
& 247 & 281
& 283 & 304
& 300
& 269
\\
Pd	 
& $b$ 
& 3.945 & 3.917
& 3.884 & 3.879
& 3.878
& 3.875
\\
& $E_{\rm coh}$ 
& 3.790 & 2.862
& 4.337 & 3.296
& 3.499
& 3.918
\\
& $B$ 
& 170 & 170
& 202 & 190
& 196
& 195
\\
Ag	 
& $b$ 
& 4.165 & 4.166
& 4.075 & 4.106
& 4.099
& 4.056
\\
& $E_{\rm coh}$ 
& 2.524 & 2.322
& 2.897 & 2.635
& 2.683
& 2.972
\\
& $B$ 
& 84 & 80
& 108 & 94
& 97
& 109
\\
\hline
\end{tabular}
\end{table}

Table \ref{tab:s22vdwcorr} also includes a characterization of the vdW-DF-cx0p performance for S22 using the approximate determination of relaxation effects,
Fig.\ \ref{fig:revS22}. That is, the Table includes a characterization of binding energies (inside square brackets in the top half) and of optimal binding separations (with average deviations listed in the bottom Table part);  Further detail is included in supplementary materials Table S.XII. When focused on relaxed calculations (for S22), we find that vdW-DF-cx0p improves the  vdW-DF-cx description, especially in terms of binding separations. The performances of the strictly parameter-free vdW-DF-cx and vdW-DF-cx0p functionals are also comparable to those of the 
dispersion-corrected versions (PBE-XDM, etc). 

Taken overall, our analysis demonstrates that the
vdW-DF-cx0 design is robust towards small changes in the Fock-exchange mixing. 
While use of vdW-DF-cx0p does improve the description over the vdW-DF-cx0, 
the improvement is not dramatic. At the same time, we find that the vdW-DF-cx0p
version does have a good average choice for the exchange mixing  $a'=0.2$, one that works
for both covalent and noncovalent molecular binding. 

This choice $a'=0.2$ is, of course, already in wide use for traditional molecular investigations, for example, used in the construction of so-called optically tuned range-separated hybrids (OTRSH).\cite{OTRSH} However, we have here documented that $a'=0.2$ is also motivated and applicable in the new vdW-DF-cx0
design\cite{DFcx02017} which can therefore provide 
concurrent descriptions of general types of molecular 
interactions.

\subsection{Robustness of vdW-DF-cx0p: extended systems}

The vdW-DF-cx0 design also aims 
to be useful for systems comprising both molecules and bulk. We therefore 
include a comparison of vdW-DF-cx0 (at $a=0.25$) and vdW-DF-cx0p (at $a'=0.2$) 
performances for a few extended systems.

Table \ref{tab:solids} contrasts the PBE, PBE0, vdW-DF-cx, vdW-DF-cx0, and vdW-DF-cx0p
description of the cubic-cell lattice constant $b$, the cohesive energy $E_{\rm coh}$
and bulk modulus $B$ for a set of traditional semiconductors (and related insulators): 
C, Si, SiC, and GaAs.
The listed experimental values for $b$ and $E_{\rm coh}$ are corrected for
vibrational zero-point energy and thermal effects, as available in Ref.\ \onlinecite{Csonka09}.
Hybrids are expected to behave reasonable for the description of bulk semiconductors 
and PBE0 generally improves the description of PBE. Similarly, while the vdW-DF-cx characterization is already at the level of PBE or better, the hybrid vdW-DF-cx 
formulations provide bulk-semiconductor descriptions that 
are accurate for the structure, cohesion, and elastic properties. 
This suggests that vdW-DF-cx0p may also serve us
for parameter-free description of molecules on a semiconducting substrate.

Simple hybrids (like PBE0 or the vdW-DF-cx0 design) should not
generally be used for description of conducting systems because they
rely on the inclusion of Fock exchange and thus lack an inherent account 
of screening. Nevertheless, we include in Table \ref{tab:solids} 
a comparison of structure, cohesion-energy, and elastic-response characterizations
for a few second-row transition metals. Again, zero-point energy and thermal
corrections on the experimental numbers are included 
from Ref.\ \onlinecite{Csonka09}, 
where available. 

One of us has previously documented that the regular nonlocal-correlation functional
vdW-DF-cx is itself highly accurate for characterizations
of the thermo-physical properties of non-magnetic transition-metal 
elements;\cite{Gharaee2017} Table \ref{tab:solids} (with results
obtained here using a different code) confirms this observation. 

Table \ref{tab:solids} furthermore shows that the vdW-DF-cx0 and vdW-DF-cx0p versions 
remain usable for some properties, specifically for structure. 
Not surprisingly, the cohesive energies worsen, although 
the vdW-DF-cx0p version performs better 
in this regards than both PBE0 and vdW-DF-cx0. However, the fact that the 
structure characterizations remain accurate for these transition metals
is promising. This suggests that vdW-DF-cx0p remains at least relevant for 
descriptions of molecular adsorption subject to full relaxation, as is often necessary.\cite{LofErh16,libxcvdW} 

Finally, we note that a full discussion of the extended systems, and especially of the metals, 
requires attention to the questions of screening exchange contributions. The vdW-DF-cx0p 
is a completely unscreened hybrid and so we are presently over-estimating the
effects of long-range exchange. Improvements relative to the vdW-DF-cx0 design\cite{DFcx02017}
are motivated and could take the form of adapting the HSE\cite{hse03,hse06} 
or OTRSH\cite{OTRSH} logic to the vdW-DF framework.

\section{Summary and conclusion}

Adapting the motivation for PBE-based hybrids,\cite{Perdew96,Burke97} we have constructed 
system-specific two-legged hybrids vdW-DF-tlh based on the vdW-DF-cx coupling-constant 
variation.\cite{ScalingAnalysis} The vdW-DF-tlh constructions are related to the idea of a
perturbation-theory approach to hybrid density functionals,\cite{Gorling93,Perdew96,Ernzerhof97} 
It combines calculations of the  functional coupling-constant 
variation and MP2 results to establish the Fock mixing fraction
$a\approx 1/m$ ($m$ integer) directly.\cite{Gorling93} The two-legged 
hybrid constructions\cite{Burke97} provide a qualitative discussion of this 
strictly parameter-free perturbation-based hybrid approach.
We emphasize that the vdW-DF-tlh design is not suggested  for
pursuing broad calculations, it is for analysis only.

Our overall discussion is based on using vdW-DF-tlh for molecular problems: 
subsets of the G2-1 atomization energy,\cite{G2-97} the G2RC set of 
reaction-energies,\cite{G2RC} the G21IP set of ionization-potentials,\cite{G2RC} the S22 set of inter-molecular binding benchmarks,\cite{S22-06,S22-10} the IDISP set of intra-molecular noncovalent interactions,\cite{gmtkn55} the
Al2X6 set of aluminum dimerization energies,\cite{G2RC,gmtkn55} and the DARC set of Diels-Alder reaction energies. The IDISP,
Al2X6 and DARC energies all test effects of electron affinities
and delocalization errors. While lacking self consistency, we find that the vdW-DF-tlh is accurate. 
This builds confidence in our qualitative results:  (a) the plausible 
all-round value for a hybrid vdW-DF-cx0 design
would be $a=0.2$, close to but different from the $a=0.25$ value that was used in 
Ref.\ \onlinecite{DFcx02017}, but also, (b) there is only partial rationale for just using single, fixed Fock-exchange mixing fraction in Eq.\ (\ref{eq:cx0def}), since there is a scatter in the values that we extract from trying to use vdW-DF-tlh. 

As an interesting aside we note that the vdW-DF-tlh analysis can be used to check if, in some specific problem, a given choice of the Fock mixing is very off for the hybrid use, for example, very different from the typical recommendations for hybrids,\cite{BurkePerspective} $a=0.25$ or $a'=0.2$. This analysis can be made using our code \textsc{ppACF}
for tracking the coupling-constant variation of the vdW-DF-cx 
functional (and other semilocal- or nonlocal-correlation 
density functionals,) as evaluated for system-specific, self-consistent electron-density solutions.\cite{ScalingAnalysis}

Our analysis leads us to explore a specific version, termed vdW-DF-cx0p, of the vdW-DF-cx0 
design.\cite{DFcx02017} This is done, because unlike vdW-DF-tlh,
the vdW-DF-cx0p (with fixed mixing $a'=0.2$) can be carried to self-consistency.
We find that vdW-DF-cx0p, compared to the original vdW-DF-cx0
form (having $a=0.25$,) generally improves the description of 
molecular systems, as expected by the vdW-DF-tlh analysis. 

We suggest using the vdW-DF-cx0p version (of the vdW-DF-cx0 design\cite{DFcx02017}) 
for both covalent and noncovalent molecular system. We make this suggestion, because 
vdW-DF-cx0p can be seen as restricting 
all parameter input to formal many-body perturbation theory:
Even the assessment of an optimal average mixing value 
$a'=0.2$ comes from analysis of the vdW-DF-tlh 
constructions (which, in turn is based on analysis 
of ACF behavior for the underlying strictly parameter-free 
vdW-DF-cx). We also observe that the vdW-DF-tlh analysis  
does indicate a finite scatter in relevant 
$a_{\rm sys}$ values for general hybrid 
characterizations of molecular interactions. We
see this scatter as, in part, expected from the 
long tradition in using
traditional semilocal-correlation hybrids.\cite{Perdew96,Burke97,Ernzerhof97,BurkePerspective,beckeperspective}

The vdW-DF-cx0p usefulness originates, in practice, from our present demonstration that a single, fixed value $a'=0.2$ can be used for a concurrent description of both covalent and 
noncovalent interactions in molecules. In addition, we find 
that it is useful in some extended-system 
cases. The vdW-DF-cx0 design\cite{DFcx02017} is robust and 
there is only moderate effects of changing the Fock-exchange mixing. 

It is interesting that our present vdW-DF-tlh-based 
identification of an optimal average mixing 
value, $a'=0.2$, for the vdW-DF-cx0 design\cite{DFcx02017} is identical to the value Becke originally extracted for standard semilocal-correlation hybrids.\cite{Becke93}  
The $a'=0.2$ value is in broad usage for molecular 
problems, for example, in the definition the OTRSH\cite{OTRSH}
that can reliably track electronic excitations. Becke's identification of $a'$ is based on 
a fit, comparing results from a range of potential  semilocal-correlation hybrids to mostly 
covalent interaction properties of molecules 
(and some atom problems). In contrast, the 
present specification is based on a formal ACF-based analysis\cite{Levy85,Gorling93,Perdew96,Burke97,ScalingAnalysis} for the underlying regular vdW-DF-cx functional. 

We view the fact that there are two independent but coinciding specifications of 
$a'=0.2$ as an indication of a soundness in the  logic of nonlocal-correlation hybrid 
functionals. The vdW-DF-cx can formally be seen as a systematic extension of semilocal 
GGA functionals,\cite{hybesc14,Berland_2015:van_waals} with a  seamless integration.
The vdW-DF-cx construction secures a highly reliable traditional-materials description 
in cases with a dense electron distribution.\cite{bearcoleluscthhy14,ErhHylLin15,LinErh16,Gharaee2017}  Such cases 
include covalent bonding in molecules. It is therefore possible to view the present 
work both as a formal rediscovery of the Becke $a'=0.2$ mixing value and as a 
demonstration that it extends to noncovalent molecular binding as well. 
The demonstration is important because vdW-DF-cx0 has a different design 
logic than the traditional semilocal-correlation 
hybrids. 

\section*{\label{sec:suppmat} Supplementary materials}

This paper has supplementary materials containing 12 tables,
denoted S.I through S.XII as individually introduced and 
referenced in the text above.  The supplementary-materials 
tables detail the two-legged hybrid constructions as well as the 
vdW-DF-cx, vdW-DF-cx0, vdW-DF-cx0p, PBE0 performance for individual 
molecular systems or reactions in the G2-1, G2RC, G21IP, S22, IDISP, 
Al2X6, and DARC benchmark sets. As such they provide they basis for 
our assessment of an optimum average mixing ratio $\langle a_{\rm sys}\rangle$
and for the performance comparisons that we summarize in tables I though VII 
(and in figures 4 and 5).

\section*{\label{sec:ack} Acknowledgement}

We thank Kristian Berland, Jung-Hoon Lee, Tonatiuh Rangel, Zhenfei Liu, and Jeffrey B. Neaton
for discussions. Work supported by the Swedish Research Council (VR) 
through grants No. 2014-4310 and 2014-5289 and by the Chalmers Area-of-Advance-Materials 
theory activity. 
Computational resources were provided by the Swedish National Infrastructure for Computing (SNIC), 
under contract 2016-10-12, and by the Chalmers Centre for Computing, Science and 
Engineering (C3SE), under contracts C3SE 605/16-4 and C3SE2018-1-10.

%\bibliographystyle{unsrt}
%\bibliography{references,hybvdw,newreferences}


\end{document}